\documentclass[twocolumn,showpacs,preprintnumbers,amsmath,amssymb,nofootinbib,superscriptaddress,english]{revtex4-1}
\usepackage{graphicx}
\usepackage{dcolumn}
\usepackage{bm}
\usepackage{epsfig}
\usepackage{graphicx}
\usepackage{hyperref}
\usepackage[usenames]{color}
\usepackage{url}
\usepackage{epstopdf}

\hypersetup{
    colorlinks=true,
    linkcolor=green,
    citecolor=blue,
}

\begin{document}

\title{Linear Growth of Structure in the Symmetron Model}

\author{Philippe~Brax}
\email[Email address: ]{philippe.brax@cea.fr}
\affiliation{Institut de Physique Theorique, CEA, IPhT, CNRS, URA 2306, F-91191Gif/Yvette Cedex, France}
\author{Carsten~van~de~Bruck}
\email[Email address: ]{c.vandebruck@sheffield.ac.uk}
\affiliation{School of Mathematics and Statistics, University of Sheffield, Hounsfield Road, Sheffield S3 7RH, UK}
\author{Anne-Christine~Davis}
\email[Email address: ]{a.c.davis@damtp.cam.ac.uk}
\affiliation{DAMTP, Centre for Mathematical Sciences, University of Cambridge, Wilberforce Road, Cambridge CB3 0WA, UK}
\author{Baojiu~Li}
\email[Email address: ]{b.li@damtp.cam.ac.uk}
\affiliation{DAMTP, Centre for Mathematical Sciences, University of Cambridge, Wilberforce Road, Cambridge CB3 0WA, UK}
\affiliation{Kavli Institute for Cosmology Cambridge, Madingley Road, Cambridge CB3 0HA, UK}
\author{Benoit~Schmauch}
\email[Email address: ]{schmauchmeister@gmail.com}
\affiliation{Institut de Physique Theorique, CEA, IPhT, CNRS, URA 2306, F-91191Gif/Yvette Cedex, France}
\author{Douglas~J.~Shaw}
\email[Email address: ]{djshaw@gmail.com}
\affiliation{DAMTP, Centre for Mathematical Sciences, University of Cambridge, Wilberforce Road, Cambridge CB3 0WA, UK}

\date{\today}

\begin{abstract}
In the symmetron mechanism, the fifth force mediated by a coupled scalar field (the symmetron) is suppressed in high-density regions due to the restoration of symmetry in the symmetron potential. In this paper we study the background cosmology and large scale structure formation in the linear perturbation regime of the symmetron model.  Analytic solutions to the symmetron in the cosmological background are found, which agree well with numerical results. We discuss the effect of the symmetron perturbation on the growth of matter perturbation, in particular the implications of the brief period of tachyonic instability caused by the negative mass-squared of the symmetron during symmetry breaking. Our analysis and numerical results show that this instability has only very small effects on the growth of structures on sub-horizon scales, and even at horizon scales its influence is not as drastic as naively expected. The symmetron fifth force in the non-tachyonic regime does affect the formation of structure in a nontrivial way which could be cosmologically observable.

\end{abstract}

\maketitle

\section{Introduction}

Scalar fields coupled to matter are generic predictions of many theories of high energy physics. In recent years, this idea has attracted a lot of attention in the context of  dark energy \cite{cst2006}, which is believed  to be a scalar field or one of its variants \cite{wcos2000,ams2000,Wetterich:1987fm,a2000,pb1999}. However, it is well known that if a scalar field couples to matter or curvature then a scalar fifth force and a modification to the gravitational law could result. Such new physics has been strongly constrained by local gravity experiments and solar system tests, so that fifth forces and modifications of  gravity must been either very short-ranged or very weak, or both. If this is true, then the effect of the scalar field is mainly to drive the accelerating expansion of the Universe, such as in the quintessence model \cite{wcos2000}.

More cosmologically interesting models could be built if the fifth force or modification to the standard Einstein gravity is only weak and/or short ranged where local experiments are performed, such as in our Solar system where matter density is high and gravity is strong, but could become strong (of gravitational strength) and long-ranged elsewhere. Over the past few years, several models have been proposed to realise this, including the chameleon model \cite{kw2004a,kw2004b,Brax:2004qh, ms2006,ms2007,lb2007,bbds2008}, the environmentally-dependent dilaton model \cite{bbds2010,bbdls2011}, the DGP model \cite{dgp2000}, the galileon model \cite{nrt2009,dev2009,dde2009,dt2010} and the symmetron model \cite{hk2010,hklm2011,wmld2011}.

Although all these models predict strong environmental dependence of the fifth force, they work in very different ways. In the chameleon model, for example, the fifth force is suppressed exponentially in high density regions where the scalar field acquires a heavy mass via its coupling to matter. The environmentally dependent dilaton model, on the other hand, drives the scalar field to some critical value $\phi_c$ in high density regions, which corresponds to a vanishing coupling and therefore vanishing fifth force (though the mass of the scalar field depends on the local matter density as well in this model).

The symmetron model, which is the topic of this work, relies on a similar mechanism to suppress the fifth force in high density regions. Here, when the matter density is high enough, the effective potential of the scalar field has a global minimum at the origin, which corresponds to a vanishing coupling to matter. When matter density drops below some critical value, the symmetry in the effective potential is broken and two local minima develop and move away from the origin, corresponding to a nonzero coupling to matter and thus a non-vanishing fifth force. If later the matter density inside a region becomes high again due to the structure formation process (such as in galaxies and galaxy clusters), the symmetry of the effective potential could be restored and the fifth force vanishes again for that region.

Local experiments could constrain the parameters of the symmetron model. Interestingly, in this constrained parameter space, there are still models which could deviate from the standard $\Lambda$CDM cosmology on large scales. The cosmological observables, in particular those relevant for the large scale structure formation, can thus provide valuable information about the symmetron, such as whether it exists, what its observational signatures are, how to differentiate it from other models, etc. In this work, we shall concentrate on the structure formation of the symmetron model in the linear perturbation regime, and show how the symmetron can affect the large scale structure of the Universe.

This paper is arranged as follows: we first briefly overview the symmetron model and how the local tests constrain its parameters in Sects.~\ref{subsect:model} and \ref{subsect:localtest}, followed by analytical and numerical study of the background cosmology in the model (Sect.~\ref{subsect:background}) and a short description of the tachyonic period, during which the scalar field has a negative mass-squared (Sect.~\ref{subsect:techyonic}). Then in Sect.~\ref{sect:linpert_general} we study the general behaviour of the linear perturbations in this model, where we show that the fifth force essentially enhances gravity within its range (given by the comoving Compton length of the symmetron field) and keeps standard gravity unmodified beyond that range. We also show that the tachyonic instability does not affect the structure formation significantly on sub-horizon scales. In Sect.~\ref{sect:linpert_numeric} we give some numerical results of the large scale structure formation and show that significant deviations from the $\Lambda$CDM paradigm can be found, which make the model cosmological interesting. We summarise and conclude in Sect.~\ref{sect:con}. Throughout this paper we shall adopt $c=\hbar=1$ and the metric convention $(-,+,+,+)$.

\section{Symmetron Dynamics}

\label{sect:model}

\subsection{The symmetron}

\label{subsect:model}

Scalar-tensor theories are characterised by their coupling to
matter and their interaction potential. In such a context, the
symmetron model was proposed in \cite{hk2010,hklm2011} and is
described by the action:
\begin{eqnarray}
S = \int \sqrt{-g} d^4 x\left[\frac{R}{2\kappa} -
\frac{1}{2}(\nabla \phi)^2 - V(\phi)\right] + S_{\rm m},
\end{eqnarray}
where $\phi$ is the scalar field and $V(\phi)$ its potential,
$S_{\rm m}\equiv S_{\rm m}\left[\Psi^{i}, \tilde{g}_{\mu
\nu}\right]$ the matter action with $\Psi^{i}$ the matter fields
which are \emph{minimally} coupled to the Jordan frame metric
$\tilde{g}_{\mu \nu} =  A^2(\phi)g_{\mu\nu}$; $g_{\mu\nu}$ is the
Einstein frame metric, which is used to compute the Ricci scalar
$R$; $\kappa=8\pi G=m_{\rm pl}^{-2}$ where $G$ is Newton's constant and
$m_{\rm pl}$ the reduced Planck mass.

For the symmetron model, the interaction potential and the
coupling function are simply chosen such that
\begin{eqnarray}
V(\phi) &=& V_{0}-\frac{1}{2}\mu^2 \phi^2 + \frac{1}{4}\lambda
\phi^4,\nonumber\\
A(\phi) &=& 1 + \frac{\phi^2}{2M^2},
\end{eqnarray}
in which $V_{0}$ is a cosmological constant and has mass dimension
4, $\mu$ and $M$ have mass dimension one and $\lambda$ is a
dimensionless model parameter. We shall see that $V_0$ is needed
to explain the recently observed accelerating expansion of the
Universe, but being a constant it bears no influence on the
dynamics of the symmetron field. Note that the potential has two
nontrivial minima while the coupling function is monotonously
increasing.

The field equations are obtained by varying the action $S$ with
respect to the symmetron field $\phi$, and we have
\begin{eqnarray}
\square\phi &=& V_{,\phi}(\phi)-A_{,\phi}(\phi)A^3(\phi)\tilde{T},\nonumber\\
\tilde{T} &=& \tilde{g}_{\mu\nu}\tilde{T}^{\mu\nu},\\
\tilde{T}^{\mu\nu} &=& \frac{2}{\sqrt{-\tilde{g}}}\frac{\delta
S_{\rm m}}{\delta\tilde{g}_{\mu\nu}},
\end{eqnarray}
where we have defined the Jordan frame energy momentum tensor
$\tilde T_{\mu\nu}$ that is related to the Einstein frame one by
$T^{\mu}{}_{\nu} = A^3(\phi) \tilde{T}^{\mu \rho}\tilde{g}_{\rho
\nu}$. Note that we raise and lower the indices of $T^{\mu}_{\nu}$
using the Einstein frame metric, $g_{\mu \nu}$. With this
definition, the field equation for $\phi$ and the Einstein
equations  become
\begin{eqnarray}
\square\phi &=& V_{\mathrm{eff},\phi}(\phi;T),\\
R^{\mu \nu} - \frac{1}{2}R g^{\mu \nu} &=& \kappa T^{\mu \nu}_{\rm tot}, \\
\end{eqnarray}
in which the symmetron field is governed by an effective potential
\begin{eqnarray}
V_{{\rm eff}}(\phi;T) &\equiv& V(\phi)-A(\phi)T,
\end{eqnarray}
and the total energy momentum is given by
\begin{eqnarray}
T^{\mu \nu}_{\rm tot} = A(\phi) T^{\mu \nu}  - g^{\mu \nu} V(\phi)
+ \nabla^{\mu}\phi \nabla^{\nu}\phi - \frac{1}{2}g^{\mu \nu}
(\nabla \phi)^2,
\end{eqnarray}
which satisfies the following conservation equation
\begin{eqnarray}
\nabla_{\mu}T^{\mu\nu} &=& \frac{A_{,\phi}}{A}\left(Tg^{\mu\nu} -
T^{\mu\nu}\right)\nabla_{\mu}\phi.
\end{eqnarray}
Note that this implies that for pressureless matter with
$T^{\mu\nu} = \rho u^{\mu}u^{\nu}$ with $u^{\mu}$ a 4-velocity
($u^{\mu}u_{\nu} = -1$), we have $-T = \rho_{\rm m}$ and
$\rho_{\rm m}$ is conserved independently of $\phi$:
\begin{eqnarray}
\nabla_{\mu}(\rho_{\rm m}u^{\mu}) = 0.
\end{eqnarray}
As is the usual practice, we define the effective mass of the
symmetron field by
\begin{eqnarray}
m^2(\phi) &=& V_{{\rm eff},\phi\phi}(\phi;T)\nonumber\\
&=& -\mu^2 + 3\lambda\phi^2 +\frac{\rho_{\rm m}}{M^2},
\end{eqnarray}
where we have used the fact that, from the above equations, the
effective potential can be rewritten as
\begin{eqnarray}\label{pot}
V_{\rm eff}(\phi) &=& \frac{1}{2}\Bigl(\frac{\rho_{\rm
m}}{M^2}-\mu^2\Bigr)\phi^2+\frac{1}{4}\lambda\phi^4.
\end{eqnarray}
Hence, in the symmetron model, as long as $\rho_{\rm m}$ is high
enough, namely $\rho_{\rm m}\ge\rho_{\star}$ where
\begin{eqnarray}
\rho_\star &\equiv& \mu^2 M^2,
\end{eqnarray}
the minimum of the effective potential is at the origin
($\phi=0$). In contrast, in vacuum, the symmetry is broken and the
potential has two nonzero minima:
\begin{eqnarray}
\phi_\star &=& \pm\frac{\mu}{\sqrt\lambda}.
\end{eqnarray}
As long as $\phi \ll M$ the effective coupling to matter reads
\begin{eqnarray}
\alpha_\phi(\phi)\ =\ m_{\rm pl}\left[\ln A\right]_{,\phi}\
\approx\ \frac{m_{\rm pl}\phi}{M^2},
\end{eqnarray}
leading to the absence of modification of gravity in dense
environments where the field vanishes. The modification of the
growth of structure depends on
\begin{equation}
\alpha \equiv 2 \alpha_\phi^2,
\end{equation}
and the order of magnitude of $\alpha$ in vacuum,
\begin{equation}
\alpha_\star = \alpha(\phi_\star),
\end{equation}
is crucial for the growth of the large scale structure. For values
of the density lower than $\rho_\star$, gravity becomes modified
in a way which could be tested cosmologically, and $\alpha_\star$
characterises the relative strength of the modification.

\subsection{Gravity tests}

\label{subsect:localtest}

In this section, we review some of the gravitational properties of the symmetron\cite{hk2010}
The symmetron model is designed to induce modifications of gravity
which could be tested in the near future, both gravitationally and
cosmologically. Requiring that the energy density at which the
curvature at the origin of the potential changes sign (roughly
when gravitation undergoes a transition from standard to modified)
is close to the current critical energy density (which implies
that gravity is modified cosmologically recently), we have the
estimate
\begin{eqnarray}
M^2 \mu^2 &\sim& H_0^2 m_{\rm Pl}^2\nonumber
\end{eqnarray}
Moreover, the modification to gravity is detectable only if it is
comparable to (or bigger than)  standard gravity, or
equivalently the effective coupling
$\alpha_\phi\sim\mathcal{O}(1)$, which implies
\begin{eqnarray}
\frac{\phi_\star}{M} &\sim& \frac{1}{\sqrt A_2},\nonumber
\end{eqnarray}
where we have defined $A_2\equiv m_{\rm Pl}^2/M^2$. These
determine the vacuum mass
\begin{eqnarray}
m^2(\phi_0)\ =\ 2\mu^2\ \sim\
\mathcal{O}\left(\frac{m_{Pl}^2}{M^2}\right)H_0^2\ =\
\mathcal{O}(A_2)H^2_0
\end{eqnarray}
and correspondingly set the interaction range of the symmetron to
be $\sim\mathcal{O}\left(m^{-1}(\phi_0)\right)$. This also
determines the self coupling \cite{hk2010}
\begin{eqnarray}
\lambda\ \sim\ \frac{\mu^2m_{pl}^2}{M^4}\ \sim\
\frac{H_0^2m_{Pl}^4}{M^6}
\end{eqnarray}
It is then crucial to have an estimate for $M$. This follows from the study of solar system tests.

Let us consider a spherical object of density $\rho$ and radius $R$. The static solutions for the field profile are obtained solving
\begin{equation} \dfrac{d^2\phi}{dr^2}+\frac{2}{r}\dfrac{d\phi}{dr}=\Bigl(\frac{\rho}{M^2}-\mu^2\Bigr)\phi +\lambda\phi^3 \end{equation}
It is convenient to simplify this equation by separating two regions with different behaviours \cite{hk2010}
\begin{eqnarray}
V_{\rm eff}(r<R) &=& \Bigl(\frac{\rho}{M^2}-\mu^2\Bigr)\frac{\phi^2}{2}
\end{eqnarray}
and
\begin{eqnarray}
V_{\rm eff}(r>R) &=& \mu^2(\phi-\phi_\star)^2
\end{eqnarray}
where one assumes that the density vanishes at infinity and outside the body.
The solutions read
\begin{eqnarray}
\phi(r<R)\ =\ \phi_{in}(r)\ =\ A\frac{R}{r}\sinh\Bigl(r\sqrt{\frac{\rho}{M^2}-\mu^2}\Bigr)
\end{eqnarray}
and
\begin{eqnarray}
\phi(r>R)\ =\ \phi_{out}(r)\ =\ B\frac{R}{r}e^{-\sqrt{2}\mu r}+\phi_\star
\end{eqnarray}
Defining the modified Newton potential at the body's surface
\begin{eqnarray}
\tilde\alpha\ \equiv\ \frac{\rho R^2}{M^2}\ =\ 6 A_2 \Phi_N
\end{eqnarray}
and the ratio of the size of the sphere to the range of the symmetron interaction
\begin{eqnarray}
\tilde \beta\ =\ \mu R
\end{eqnarray}
we find that
\begin{equation}
A=\frac{(1+\sqrt{2}\tilde \beta)\phi_\star}{\sqrt{\tilde \alpha-\tilde \beta^2}\cosh\sqrt{\tilde \alpha-\tilde \beta^2}+\sqrt{2}\tilde\beta\sinh\sqrt{\tilde \alpha-\tilde \beta^2}}
\end{equation}
Notice that $\tilde \alpha \gg \tilde \beta^2$ as long as $ \rho \gg \rho_\star$.
There are two types of solutions depending on the values of $\tilde \alpha$. When $\tilde\alpha\ll 1$ we have
\begin{eqnarray}
A &\approx& \frac{\phi_\star}{\sqrt{\tilde\alpha}}\left(1-\frac{\tilde\alpha}{2}\right) \\
B &\approx& -\frac{\tilde\alpha\phi_\star}{3}
\end{eqnarray}
The scalar force acting on a test mass outside the sphere is
\begin{eqnarray}
\frac{F_{\phi}}{\rm mass}=\frac{\tilde \alpha_{\phi}}{m_{Pl}}\partial^i\phi\sim\frac{\phi_\star^2}{M^2}\frac{\tilde\alpha}{3}\frac{R}{r^2}
\end{eqnarray}
when the Newtonian force is $\frac{F_N}{\rm mass}=\frac{\rho R^3}{6m_{Pl}^2r^2}$ implying that
\begin{equation}
\frac{F_{\phi}}{F_N}\sim 2\tilde\alpha\frac{\phi_\star^2}{M^2}\frac{m_{Pl}^2}{\rho R^2}\sim2\tilde\alpha\frac{M^2}{\rho R^2}=\mathcal{O}(1).
\end{equation}
This implies that the scalar force is not screened. On the other hand when $\tilde \alpha \gg 1$ we have
\begin{eqnarray}
A &\approx& \frac{2\phi_0}{\sqrt{\tilde\alpha}}e^{-\sqrt{\tilde\alpha}} \\
B &\approx& -\phi_0\left(1-\frac{1}{\tilde\alpha}\right)
\end{eqnarray}
The scalar force is now
\begin{eqnarray}
F_{\phi}=\frac{\alpha_{\phi}}{m_{Pl}}\partial^i\phi\sim\frac{\phi_\star^2}{M^2}\frac{R}{r^2},
\end{eqnarray}
hence
\begin{equation}
\frac{F_{\phi}}{F_N}\sim\frac{\phi_0^2}{M^2}\frac{m_{Pl}^2}{\rho R^2}\sim\frac{1}{\tilde\alpha}\ll 1,
\end{equation}
leading to a large screening of the scalar force.

Phenomenologically, one must impose that in our galaxy $\tilde \alpha_G\gg 1$.
Imposing at least $\tilde \alpha_G\gtrsim  10$ and upon using  $\Phi_G\sim 10^{-6}$, one gets  $M\lesssim 10^{-3}m_{Pl}$ or equivalently
\begin{equation}
A_2 \gtrsim 10^6.
\end{equation}
This implies that $\frac{\phi_\star }{M}\lesssim 10^{-3}$.


The range of the symmetron in vacuum is given by
\begin{eqnarray}
\mu^{-1}\lesssim 10^3H_0^{-1}\sim \ 1\ \rm  Mpc
\end{eqnarray}
which corresponds to relevant scales for astrophysics. We will come back to this point later.
If $M\approx10^{-3}m_{Pl}$ then the scalar field is just about screened by the sun as $\Phi_{\odot}\sim 10^{-6}\Rightarrow\alpha_{\odot}\approx 10$. On the other hand, the earth is not screened as $\Phi_{\oplus}\sim 10^{-9}$ and  $\alpha_{\oplus}\approx 10^{-2}$.
What matters then for solar system tests is the value of the field in the galaxy:
\begin{equation}
\frac{\phi_G}{M}\approx\frac{M}{m_{Pl}}\frac{R_G}{\sqrt{\alpha_G}R_{s}}\exp\left(-\frac{R_G-R_s}{R_G}\sqrt{\alpha_G}\right).
\end{equation}
The most stringent constraint in the solar system is the Cassini bound $\vert \gamma -1\vert \lesssim 10^{-5}$ on the Eddington parameter $\gamma$ \cite{Bertotti:2003rm}.
In the Einstein frame, the metric is expressed as
\begin{equation}
g_{00}=-(1+2\Phi_E)
\end{equation}
\begin{equation}
g_{ij}=(1-2\Phi_E)\delta_{ij}
\end{equation}
while the one that particles follow is the Jordan frame metric
\begin{equation}
\tilde{g}_{00}=-(1+2\Phi_J)
\end{equation}
\begin{equation}
\tilde{g}_{ij}=(1-2\gamma\Phi_J)\delta_{ij},
\end{equation}
from which we have \cite{hk2010}
\begin{equation}
\gamma-1\approx-\frac{\frac{\phi^2}{M^2}}{\frac{\phi^2}{2M^2}+\Phi_E}\approx-\frac{\phi^2}{M^2\Phi}.
\end{equation}
Denoting by $z_\star$ the moment when the symmetron potential becomes unstable at the origin, we have
\begin{equation}
\mu^2M^2=3H_0^2m_{Pl}^2\Omega_{m0}(1+z_*)^3.
\end{equation}
Fixing for instance
$\alpha_G\approx 600$ and  $M\approx 10^{-4}m_{Pl}$ while using for the galaxy $\tilde\beta\sim4\times10^{-2}(1+z_*)^{\frac{3}{2}}$
we get a bound on admissible $z_\star \lesssim 20$ to evade the Cassini bound $\vert \gamma -1 \vert \le 10^{-5}$.


\subsection{Cosmological evolution}

\label{subsect:background}

%

We are interested in the symmetron evolution during the
matter dominated era, when the symmetron potential is negligible
compared to the matter density $\rho_{\rm m}$. This can be seen by
evaluating the value of the potential at its absolute minimum in
vacuum $\phi_\star$ as the potential at the effective minimum is
monotonously decreasing with $\rho_{\rm m}$
\begin{eqnarray}
V(\phi_\star) &=& -\frac{\mu^4}{4\lambda}\nonumber
\end{eqnarray}
which is always $\lesssim10^{-6}\ H_0^2m_{Pl}^2$. Moreover, as
$\phi_\star /M \lesssim 10^{-3}$, the coupling function $A(\phi)$
is also approximately equal to one. The Friedmann equation is then
to a high precision
\begin{eqnarray}
\label{H} H^2 &=& \frac{\rho_{\rm m}}{3m_{Pl}^2}
\end{eqnarray}
in the matter dominated era.

\begin{figure*}[tbh]
\begin{center}
\includegraphics[width=9cm]{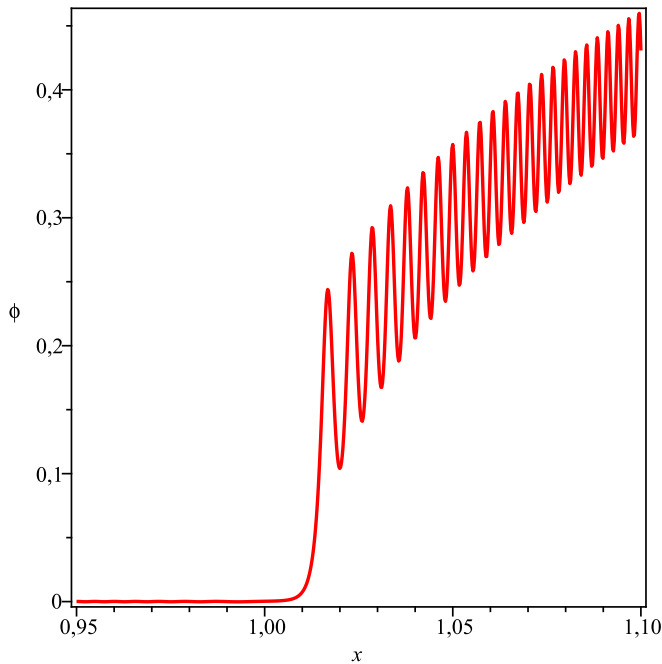}
\caption[]{The time evolution of the background symmetron field $\varphi=\phi/\phi_\star$, in which the parameters are chosen as $\kappa=4\times10^3$, $z_\star\equiv z(t_\star)=3$ and $\alpha_\star =1$. Note that immediately after the symmetry breaking, i.e., $x=1$, $\varphi$ grows exponentially but then quickly begins to oscillate. The oscillation is around the minimum of the effective symmetron potential $V_{\rm eff}(\phi)$. Note also that $-\varphi$ is also a possible branch of solution but it gives nothing new.
\label{fig:sym_osi}}
\end{center}
\end{figure*}

A full picture of the time evolution of the reduced symmetron field $\varphi\equiv\phi/\phi_\star$ is shown in Fig.~\ref{fig:sym_osi}, and we can see that after the symmetron breaking at $x(\equiv t/t_\star)=1$ the symmetron deviates from zero, and finally starts to oscillate around the moving minimum of the effective potential ($V_{\rm eff}$). The detailed evolution of $\varphi$ is of course more complicated, but still understandable analytically in certain limits, as we shall show now.

Let us denote by $ t_\star$ the instant when $\rho_{\rm
m}=\rho_\star$ and $a_\star$ the corresponding scale factor. The
scalar field equation of motion reads now
\begin{eqnarray}
\ddot{\phi}
+\frac{2}{t}\dot{\phi}+\mu^2\Bigl(\frac{t_\star^2}{t^2}-1\Bigr)\phi+\lambda\phi^3=0
\end{eqnarray}
which can be cast in a dimensionless fashion by defining $\kappa=\mu t_\star$ in addition to the $\varphi$ and $x$ defined above,
where we have the limit
\begin{eqnarray}
\kappa^2\ =\ \mu^2 t_\star^2\ =\ \frac{4\mu^2}{9H_\star^2}\ =\
\frac{4m_{Pl}^2}{3M^2}\ =\ \frac{4}{ 3} A_2\ \gtrsim\ 10^6
\end{eqnarray}
The dimensionless version of the symmetron equation of motion is
then given by
\begin{equation}\label{fadim}
\varphi''(x)+\frac{2}{x}\varphi'(x)+\kappa^2\Bigl(\frac{1}{x^2}-1\Bigr)\varphi(x)+\kappa^2\varphi(x)^3
= 0,
\end{equation}
where $'\equiv d/dx$.

\begin{figure*}[tbh]
\begin{center}
\includegraphics[width=7.1cm]{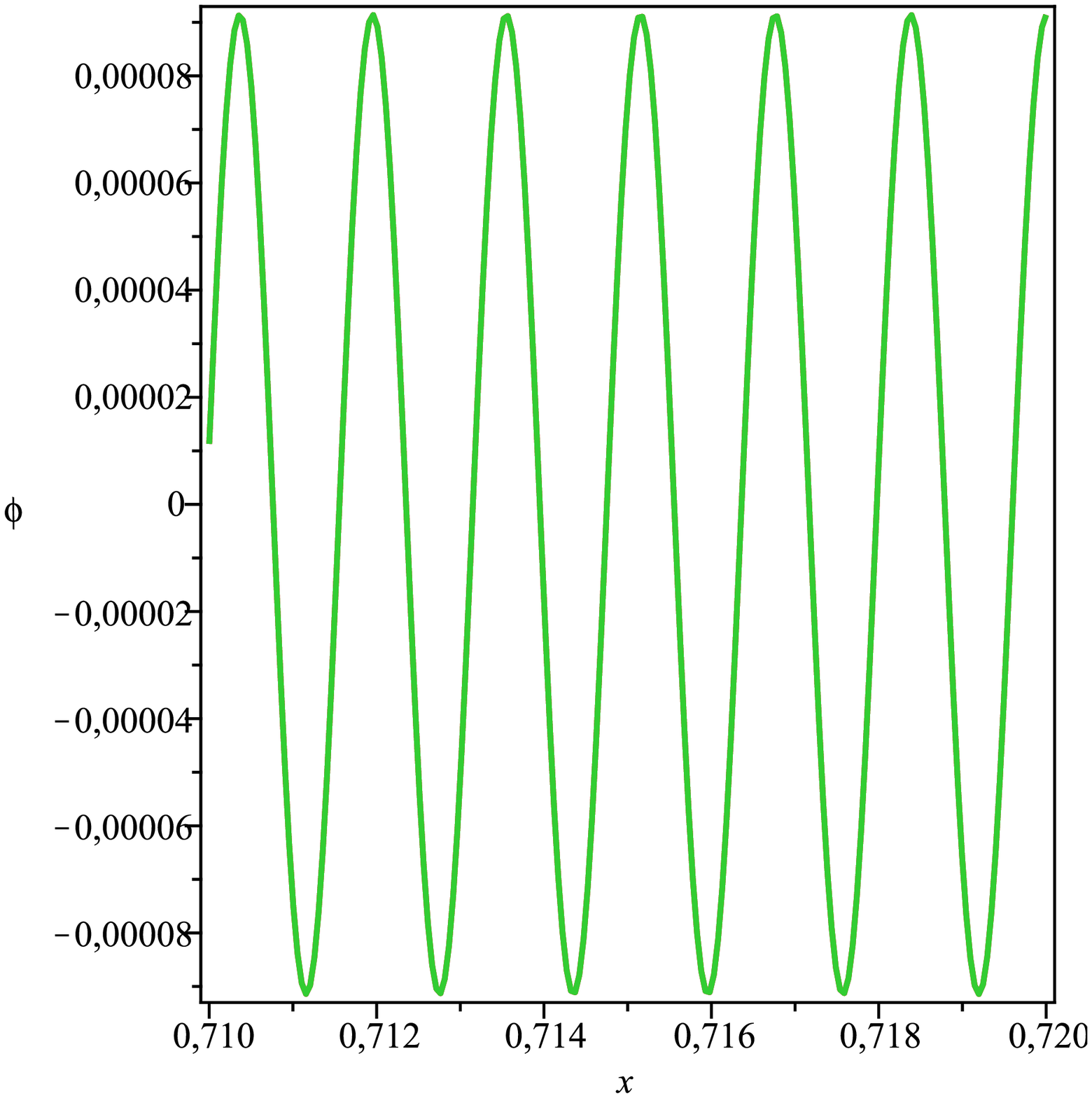}
\includegraphics[width=7.1cm]{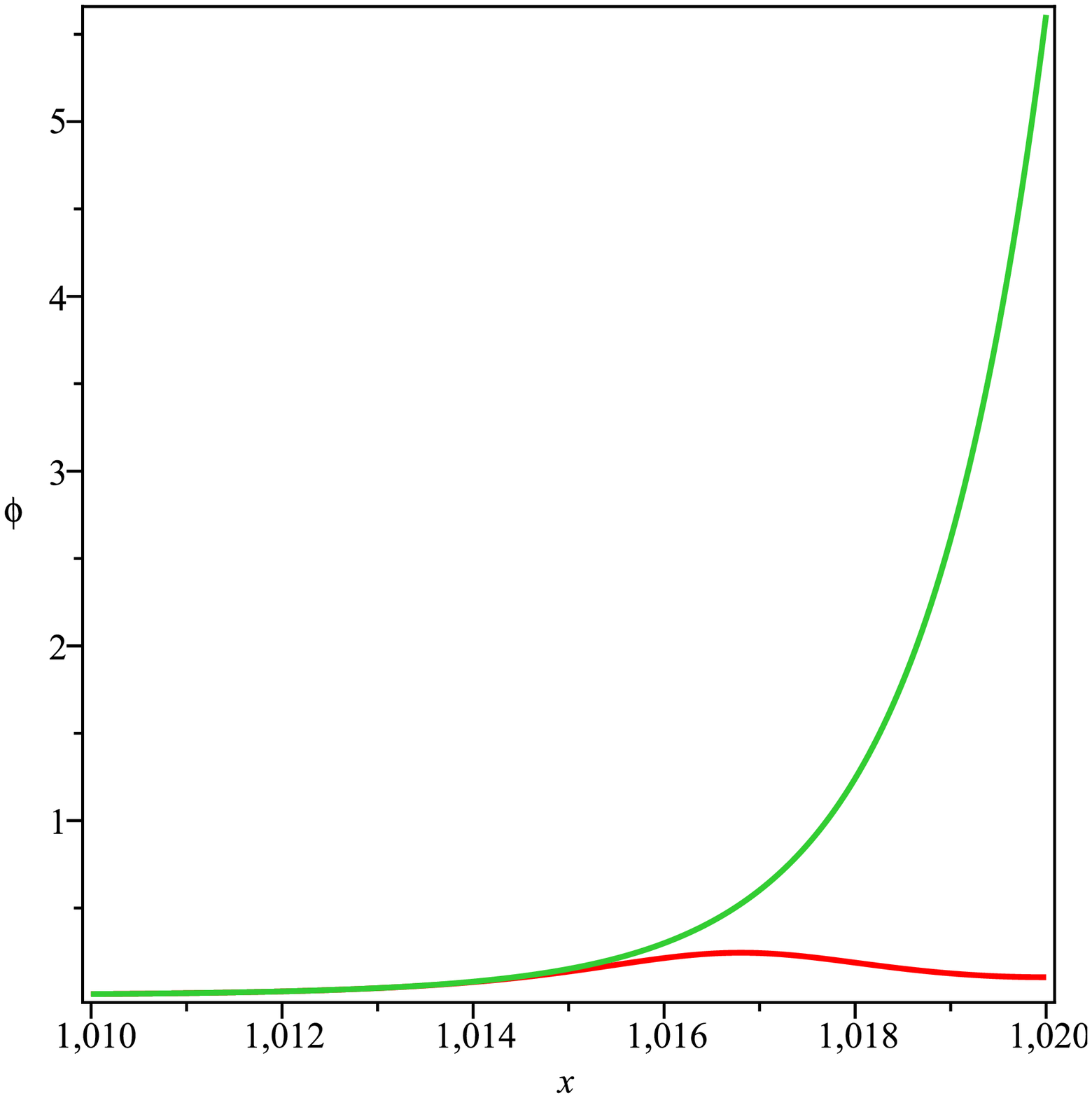}
\includegraphics[width=7.1cm]{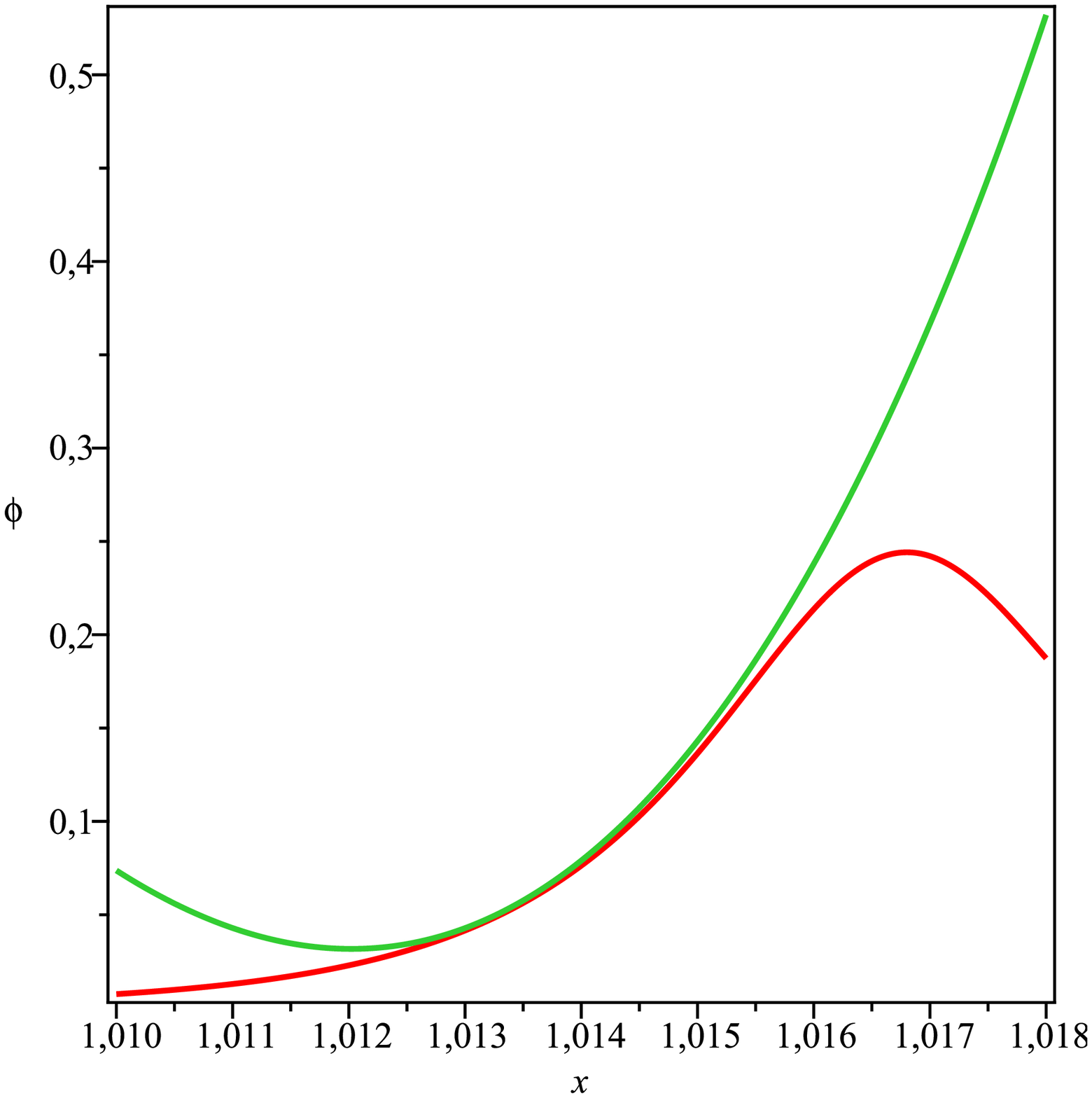}
\includegraphics[width=7.1cm]{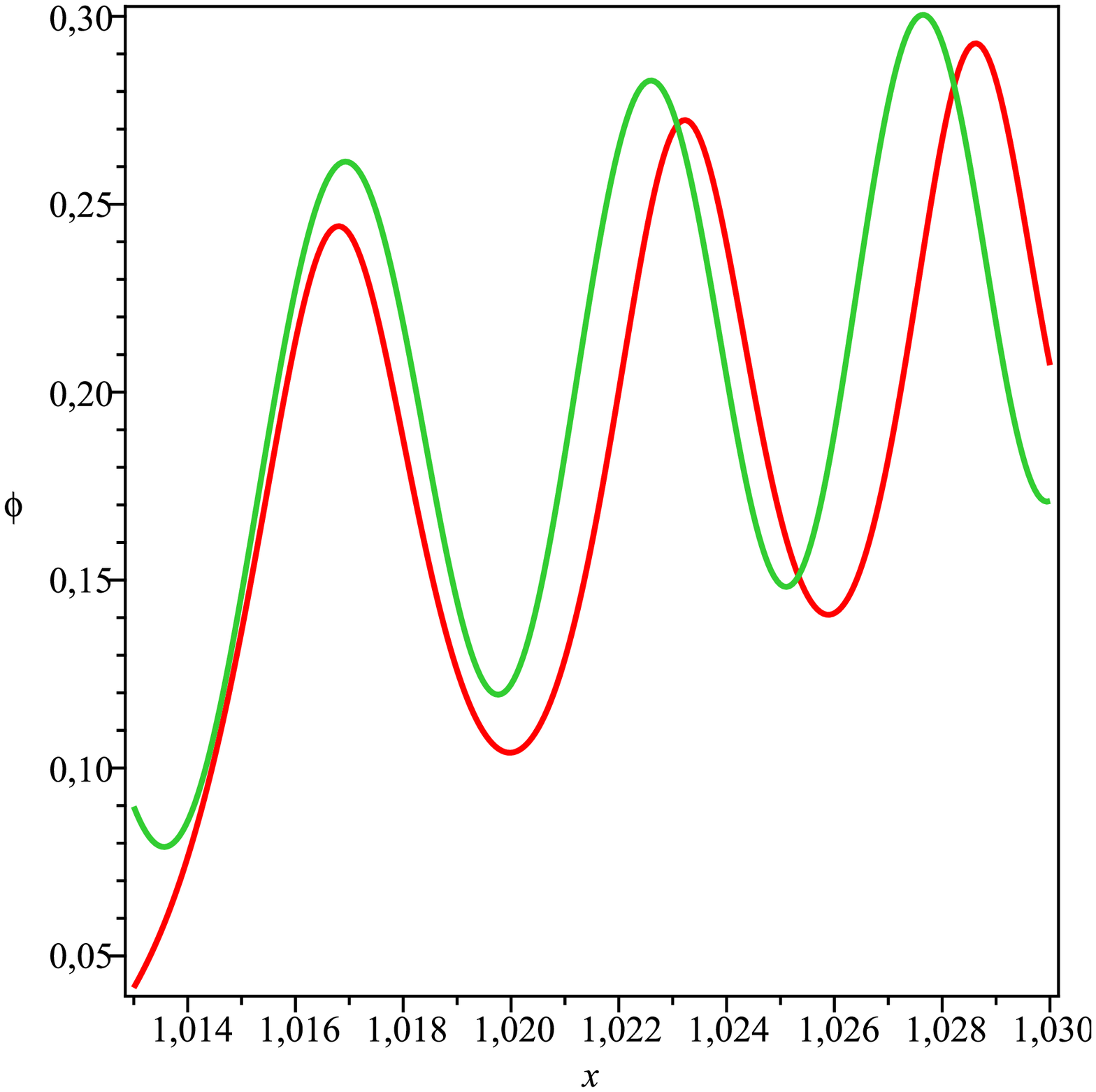}
\caption[]{The numerical results and analytic approximations for the background evolution of $\varphi=\phi/\phi_\star$ in different stages, for which the model parameters are chosen as $\kappa=4\times10^3$, $z_\star\equiv z(t_\star)=3$ and $\alpha_\star =1$. The initial condition for the symmetron is chosen as $\varphi(z_{eq})=10^{-2}$. {\it Upper left panel}: the numerical oscillation solution to $\varphi$ before the symmetry breaking, i.e., when $x<1$ or equally $t<t_\star$. {\it Upper right panel}: the numerical (red curve) and analytic (green curve) solutions to $\varphi$ immediately after the symmetry breaking, when the scalar field lags behind the minimum of its effective potential $V_{\rm eff}$. {\it Lower left panel}: {The analytical (green curve) and numerical (red curve) solutions when the symmetron tracks the inflection point}. {\it Lower right panel}: the numerical (red curve) and analytic (green curve) solutions to $\varphi$ after the scalar field settles down to the minimum of $V_{\rm eff}$ and starts to oscillates about it.
\label{fig:background}}
\end{center}
\end{figure*}

Before the curvature at the origin changes its sign, the symmetron
oscillates about the origin $\phi=0$, which is then the minimum of
the effective potential. Assuming that the initial amplitude is
small, the symmetron equation of motion simplifies to
\begin{eqnarray}\label{fadim1}
x^2\varphi''+2x\varphi'-\kappa^2(x^2-1)\varphi &\approx& 0
\end{eqnarray}
The solutions read
\begin{eqnarray}\label{bessel}
\varphi(x) &=& \frac{1}{\sqrt{x}}\left[a_1I_{i\iota}(\kappa
x)+b_1K_{i\iota}(\kappa x)\right]
\end{eqnarray}
where $\iota=\frac{\sqrt{4\kappa^2-1}}{2}\approx\kappa $,  $a_1$
and $b_2$ are constants of integration which could be determined
by the initial conditions. As long as $x\ll 1$ we can expand
\begin{eqnarray}
\phi(t) &=&
\phi_\star\sqrt{\frac{t_\star}{t}}\Bigl[\tilde{a}\cos\Bigl(\iota\ln\frac{t}{t_\star}\Bigr)+\tilde{b}\sin\Bigl(\iota\ln\frac{t}{t_\star}\Bigr)\Bigr].
\label{fond0}
\end{eqnarray}
The upper left panel of Fig.~\ref{fig:background} shows a numerical example of the evolution of $\varphi$ with respect to $x$ for $x<1$ ($t<t_\star$), from which we can see the oscillation around the global minimum of $V_{\rm eff}(\phi)$: $\phi=0$.

After the change of curvature, the field rolls away from the
origin and lags behind the minimum of the effective potential. The
minima of the effective potential are at
\begin{eqnarray}
\phi_{\pm}(t) &=&
\pm\frac{1}{\sqrt{\lambda}}\sqrt{\mu^2-\frac{\rho_{\rm
m}(t)}{M^2}}.
\end{eqnarray}
Before reaching one of the minima and oscillating around it, the
symmetron field will first linger around the origin before
following the inflection point where the curvature of the
effective potential vanishes and then settling down to that new
minimum. At the inflection point the field is close to
\begin{eqnarray}\label{eq:near_inflex}
\varphi &=& \frac{1}{\sqrt{3}}\sqrt{1-\frac{1}{x^2}}
\end{eqnarray}
where $\frac{d^2 V_{\rm eff}}{d\phi^2}=0$, and in deriving this
equation we have used the facts that $\rho_{\rm m}a^3 = \rho_\star
a^3_\star$ (conservation of matter) and that in the
matter dominated era $a\propto t^{2/3}$. At this time, the first
derivative of the symmetron effective potential becomes
\begin{eqnarray}
\frac{dV_{\rm eff}(\phi)}{d\phi} &\approx&
\mp\frac{2\mu^3}{3\sqrt{3\lambda}}\Bigl(1-\frac{t_\star^2}{t^2}\Bigr)^{\frac{3}{2}}.\nonumber
\end{eqnarray}
The symmetron equation of motion becomes then
\begin{eqnarray}
\varphi'' +\frac{2}{x}\varphi= \frac{2\kappa^2}{3\sqrt 3}
\left(1-\frac{1}{x^2}\right)^{3/2}
\end{eqnarray}
the solutions of which are
\begin{eqnarray}
\varphi(x) =
\pm\frac{2\kappa^2}{3\sqrt{3}}\Bigl[\frac{1}{x}\arctan\frac{1}{\sqrt{x^2-1}}+\frac{7}{3}\frac{\sqrt{x^2-1}}{x}\nonumber
\end{eqnarray}
\begin{eqnarray}
+\frac{1}{6}x\sqrt{x^2-1}-\frac{3}{2}\ln(x+\sqrt{x^2-1})-\frac{a_2}{x}+b_2\Bigr]
\label{fond2}
\end{eqnarray}
where $a_2$ and $b_2$ are again constants of
integration which depend on the initial conditions.  The upper right panel of Fig.~\ref{fig:background} shows the comparison of the numerical solution with the above analytic approximation (\ref{bessel}) , where we can see that for small $x$ just over 1  the agreement is very good. In the lower left panel of the same figure, one shows the short period of time when the symmetron tracks the inflection point (\ref{fond2}) before reaching the moving minimum.

In a third phase, the symmetron field catches up with the minimum
of its effective potential. Let us define
$\psi=a^{\frac{3}{2}}\phi$ which satisfies
\begin{eqnarray}\label{fadim3}
\ddot{\psi}+\mu^2\left(\frac{t_\star^2}{t^2}-1\right)\psi+\lambda\frac{\psi^3}{a^3}
&=& 0
\end{eqnarray}
Around the minimum of the effective potential,
$\psi_{+}(t)=\frac{\mu}{\sqrt{\lambda}}a^{\frac{3}{2}}\sqrt{1-\frac{t_\star^2}{t^2}}$,
one can expand to first order $\psi= \psi_+ +\psi^{(1)}$ leading
to
\begin{eqnarray}\label{ordre1}
\ddot{\psi}^{(1)}+2\mu^2\left(1-\frac{t_\star^2}{t^2}\right)\psi^{(1)} &=& -\ddot{\psi_+}\nonumber\\
&=& \frac{\mu}{\sqrt{\lambda}}\frac{a_\star^{3\over2}}{t_\star^2}\left(\frac{t^2}{t_\star^2}-1\right)^{-{3\over2}}
\end{eqnarray}
This is a forced oscillation where the forcing term is negligible for $x\gg 1$ and the mass-squared term satisfies  $m^2=2\mu^2\left(1-\frac{t_\star^2}{t^2}\right)$ The time variation of the mass compared to the period of the oscillation is given by:
$$ \frac{\dot{m}}{m^2}=\frac{1}{\sqrt{2}\kappa}\left(\frac{t^2}{t_\star^2}-1\right)^{-{3\over2}} $$
which is small when $\frac{\dot{m}}{m^2}\ll 1$, a condition satisfied when  $ \frac{t^2}{t_\star^2}-1\gg\kappa^{-{2\over3}} $, i.e., large enough $x$.  The solution for large enough $x$ is then given by
\begin{eqnarray}
\psi^{(1)}_{H}(t) &=& \left(\frac{t^2}{t^2-t_\star^2}\right)^{1\over4}\left[a_3\cos\Omega(t)
+b_3\sin\Omega(t)\right],
\end{eqnarray}
where
\begin{eqnarray}
\Omega(t) &\equiv& \sqrt{2}\kappa\left(\sqrt{\frac{t^2}{t_\star^2}-1}+\tan^{-1}\frac{t_\star}{\sqrt{t^2-t_\star^2}}\right)\nonumber
\end{eqnarray}
and $a_3$ and $b_3$ are constants of integration which depend on the initial conditions.

A better solution can be obtained by taking into account the forcing term
whose characteristic time is $\tau=\left\vert\frac{\dot{\ddot{\psi_+}}}{\ddot{\psi_+}}\right\vert\sim\frac{3}{t}\frac{1}{1-\frac{t_\star^2}{t^2}}$. The forcing term is slowly varying when $\frac{1}{m\tau}\ll1$, which reads
\begin{eqnarray}
\frac{1}{m\tau} &\sim& \frac{3}{\kappa}\frac{t_\star}{t}\left(1-\frac{t_\star^2}{t^2}\right)^{-{3\over2}}\nonumber
\end{eqnarray}
and is indeed small when the period is slowly varying. In this case, a better approximation is given by
\begin{eqnarray}
\psi^{(1)}(t) &=& -\frac{\ddot{\psi_+}}{m^2}+\psi^{(1)}_{H}
\end{eqnarray}
where the forcing term rapidly becomes negligible. The lower right panel of Fig.~\ref{fig:background} shows the analytic and numerical solutions for this stage; note the drift in the period of the oscillation due to the fact that close to $x=1$ the ratio $\dot{m}/m^2$ is not negligible.

\subsection{Tachyonic instability}

\label{subsect:techyonic}

The mass-squared of the symmetron field is briefly negative when the symmetron lags behind the minimum of its effective potential. Defining
\begin{eqnarray}
\Delta t \equiv t_{f}-t_\star,\nonumber
\end{eqnarray}
in which $t_f$ is the time when the symmetron field settles down to the effective potential minimum, one could estimate $\Delta t/t_\star$. In the case for $\kappa=10^3$, a good approximation for the time spent in the tachyonic regime can be numerically fitted as 
\begin{eqnarray}
\frac{\Delta t}{t_\star} &\approx& -2\times10^{-4}\log^2(\varphi(t_{eq}))-9.7\times10^{-3}\log(\varphi(t_{eq}))\nonumber\\
&& +2\times10^{-2}.\nonumber
\end{eqnarray}
As it stands, this tachyonic period is extremely short (see the  discussion below).
We will  see shortly that the effect of the negative mass-squared is only relevant on  large enough scales for cosmological perturbations, and even for those large scales it is quite insignificant.

\section{Linear Perturbations}

\label{sect:linpert_general}

\subsection{Growth of structure}

We will now be interested in the growth of linear perturbations in the matter dominated era, and for that we work in the Newtonian gauge where the perturbed metric in the absence of any anisotropic stress is given by
\begin{eqnarray} \label{eq:metric}
ds^2 &=& -(1+2\Phi_N)dt^2+a^2(t)(1-2\Phi_N)dx^{i}dx_{i}
\end{eqnarray}
where $\Phi_N$ represents the Newtonian potential. For simplicity let us assume that matter comprises a single fluid of pressureless particles with the energy momentum tensor
$\hat T^{\mu}_{\;\;\;\nu}=A\rho u^{\mu}u_{\nu}$, $\rho$ being the conserved energy density and  $u^{\mu}$ the four-velocity of the matter particles. In general the conservation equation reads
\begin{eqnarray}
\dfrac{d\rho}{d\tau}+3h\rho &=& 0
\end{eqnarray}
where $\tau$ is the proper time along the particle trajectories.
The local Hubble expansion rate is defined as $3h\equiv\nabla_{\mu}u^{\mu}.$
The perturbed conservation equation is then
\begin{eqnarray}\label{consm}
\dot{\delta} &=& -\theta+3\dot{\Phi}_N
\end{eqnarray}
The Euler equation in terms of the divergence $\theta\equiv\mathbf{\nabla\cdot u}$ of the velocity field becomes here
\begin{eqnarray}
\dot{\theta}+2H\theta+\frac{\Delta\Phi_N}{a^2}+\frac{\alpha_{\phi}}{m_{Pl}}\left(\frac{\Delta\delta\phi}{a^2}+\dot{\phi}\theta\right) &=& 0,
\end{eqnarray}
which is modified by the presence of the symmetron field as indicated by the last term on the left-hand side.

Similarly, the modified Poisson equation now involves the perturbation of the Einstein frame matter (and scalar field) density $\hat\rho = A(\phi)\rho$ and reads
\begin{eqnarray}\label{poisson}
\vec{\nabla}^2\Phi_N &=& \frac{1}{2m_{Pl}^2}A(\phi)\rho\left(\delta+\frac{\alpha_{\phi}}{m_{Pl}}\delta\phi\right)
\end{eqnarray}
which becomes in Fourier space when $A(\phi)\approx 1$
\begin{equation}\label{poisson} -\frac{k^2}{a^2}\Phi_N=\frac{3}{2}H^2\left(\delta+\frac{\alpha_{\phi}}{m_{Pl}}\delta\phi\right) \end{equation}
The scalar field equation of motion is expressed as
\begin{eqnarray}\label{kglin}
&&\ddot{\delta\phi}+3H\dot{\delta\phi}+\left(\frac{k^2}{a^2}+m^2(\phi)\right)\delta\phi\nonumber\\
&=& -2\Phi_N V'_{eff}(\phi)+4\dot{\Phi}_N\dot{\phi}-\frac{\bar{\rho}\phi}{M^2}\delta
\end{eqnarray}
From the above equations one can derive a second order differential equation for $\delta=\frac{\delta \rho}{\rho}$, with the coupling to the scalar field taken into account:
\begin{widetext}
\begin{eqnarray}
&&\left(1-\frac{9}{2}\frac{a^2H^2}{k^2}\right)\ddot{\delta}+\left[2H+\beta\dot{\phi}\left(1+\frac{9}{2}\frac{a^2H^2}{k^2}\right)\right]\dot{\delta}
-\left(\frac{3}{2}H^2+\frac{9}{2}\frac{a^2H^2}{k^2}\left[\beta H\dot{\phi}-\frac{1}{2}H^2\right)\right]\delta \nonumber\\
&=& -\frac{9}{2}\frac{a^2H^2}{k^2}\beta\ddot{\delta\phi}-\frac{9}{2}\frac{a^2H^2}{k^2}\left(2\dot{\beta}+\beta^2\dot{\phi}\right)\dot{\delta\phi}
  +\left[\frac{3}{2}\beta H^2-\frac{k^2}{a^2}\beta-\frac{9}{2}\frac{a^2H^2}{k^2}\left(\ddot{\beta}+\frac{1}{2}\beta H^2+\beta\dot{\beta}\dot{\phi}-\beta^2H\dot{\phi}\right)\right]\delta\phi,
\end{eqnarray}
\end{widetext}
where have defined $\beta\equiv\frac{\alpha_\phi}{m_{pl}}$.

\subsection{Adiabatic approximation}

We have seen above that the symmetron field lags behind the minimum of its effective potential for a very brief period when the mass-squared $m^2(\phi)<0$.
In such a tachyonic phase, the perturbation $\delta\phi$ grows exponentially for modes such that $\frac{k^2}{a^2} + m^2 (\phi) <0$ \cite{wmld2011} as can be seen from Eq.~(\ref{kglin}). At the time of symmetry breaking, the effective mass-squared vanishes and then decreases to a fraction of $-\mu^2$ before increasing to its value at the effective potential minimum. Clearly larger-scale modes, for which $k$ is smaller, are easily  in the tachyonic regime; as $\mu$ is much larger than the Hubble expansion rate today, some of the tachyonic modes could indeed be sub-horizon (big $k$) and this could possibly influence the growth of large-scale structure on sub-horizon scales.

Outside the tachyonic regime, the evolution equation of the symmetron perturbation is sourced by the matter perturbation. Assuming that the symmetron field tracks the effective potential minimum, i.e., for non-tachyonic modes and neglecting the short period during which the symmetron field lags behind the minimum, its equation of motion in the sub-horizon limit becomes
\begin{eqnarray}
\left(\frac{k^2}{a^2}+m^2(\phi)\right)\delta\phi &\approx& -\frac{\rho\phi}{M^2}\delta
\end{eqnarray}
where the time derivatives are much smaller than the spatial derivatives in the sub-horizon limit and thus are neglected.
From this equation the symmetron field perturbation can be solved as
\begin{eqnarray}\label{eq:deltaphi}
\delta\phi &\approx& -\frac{a^2}{k^2}\frac{\phi}{M^2}{\rho}\frac{1}{1+\frac{a^2m^2(\phi)}{k^2}}\delta,
\end{eqnarray}
which shows that the symmetron field perturbations tracks that of the matter density.
This in turn implies that matter perturbation well within the horizon grows according to
\begin{eqnarray}
\ddot{\delta}+(2H+\beta\dot{\phi})\dot{\delta}-\frac{3}{2}H^2\delta &\approx& -\frac{k^2}{a^2}\beta\,\delta\phi \nonumber
\end{eqnarray}
or equivalently
\begin{eqnarray}
\ddot{\delta}+ 2H\dot{\delta}-\frac{3}{2}H^2\left[1+\alpha (\phi)\frac{1}{1+\frac{a^2m^2(\phi)}{k^2}}\right]\delta &=& 0.
\end{eqnarray}
Therefore, in this adiabatic approximation valid for non-tachyonic modes and neglecting the interval of time when the symmetron lags behind the minimum, we find that
gravity is modified according to the comoving Compton radius $a m$ with an amplitude depending on $\alpha$. Structures on scales outside the Compton radius grow as in GR
\begin{eqnarray}
\delta &\sim& a\nonumber
\end{eqnarray}
while those on scales inside the Compton radius have a modified growth due to the renormalised Newton constant
\begin{eqnarray}
G_{\rm eff} &=& G_N (1+ \alpha).
\end{eqnarray}
We will see in the following that this result is hardly modified by the tachyonic instablity.

\subsection{Tachyonic instability}

\begin{figure}[tbh]
\begin{center}
\includegraphics[width=7.1cm]{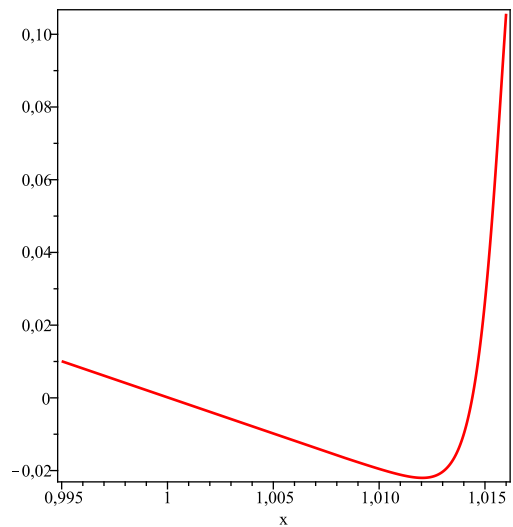}
\caption[]{The evolution of the mass-squared of the symmetron, from which we can see that for a short period immediately after the symmetry breaking ($x=1$) the mass-squared becomes negative and the symmetron field and its perturbation experience tachyonic instabilities.
\label{fig:mass}}
\end{center}
\end{figure}

As mentioned earlier, if the mass-squared of the symmetron field $m^2(\phi)$ becomes negative (see Fig.~\ref{fig:mass}), then the perturbation will undergo an unstable growth which could be problematic in some cases. We have also seen that such tachyonic instability problem is most likely to plague the large-scale (small $k$) modes. The purpose of this subsection is to assess how big the impact it could have on the growth of matter perturbations.

We are interested in modes which enter the horizon after matter-radiation equality and before the tachyonic instability happens.
Normalising $a_\star=1$, this corresponds to
\begin{eqnarray}
\frac{k}{H_\star} &\geqslant& 1
\end{eqnarray}
and
\begin{eqnarray}
\frac{k}{H_\star}\ \leqslant\ \sqrt{\frac{1+z_{eq}}{1+z_\star}}\ \lesssim\ 60
\end{eqnarray}
The tachyonic modes can be conveniently studied using the reduced symmetron perturbation
\begin{eqnarray}
\delta\varphi &\equiv& \frac{\delta\phi}{\phi_\star}\nonumber
\end{eqnarray}
and the parameters
\begin{eqnarray}
\omega &\equiv& \frac{\phi_\star}{M}\lesssim 10^{-3}\nonumber
\end{eqnarray}
and
\begin{eqnarray}
\xi_k\ \equiv\ kt_\star\ =\ \frac{2k}{3H_\star}.\nonumber
\end{eqnarray}
Keeping terms in $\omega^2$ (or $\beta$), neglecting terms in $\omega^4$ (or $\beta^2$), and considering that the variations in both $\varphi$ and $\delta\varphi$ are very rapid in the tachyonic period,  the growth equation simplifies to 
\begin{eqnarray}
\delta''+\frac{4}{3x}\delta'-\frac{2}{3x^2}\delta &=& -\frac{2\omega^2}{\xi_k^2x^{2\over3}}(\varphi\delta\varphi''+2\varphi'\delta\varphi'+\varphi''\delta\varphi)\nonumber
\end{eqnarray}
in the sub-horizon limit for the  tachyonic modes.
Notice that in the absence of the rapid variation of $\delta\varphi$ due to the tachyonic instability, this equation reduces to the
growth equation in GR.

\begin{figure*}[tbh]
\begin{center}
\includegraphics[width=7.1cm]{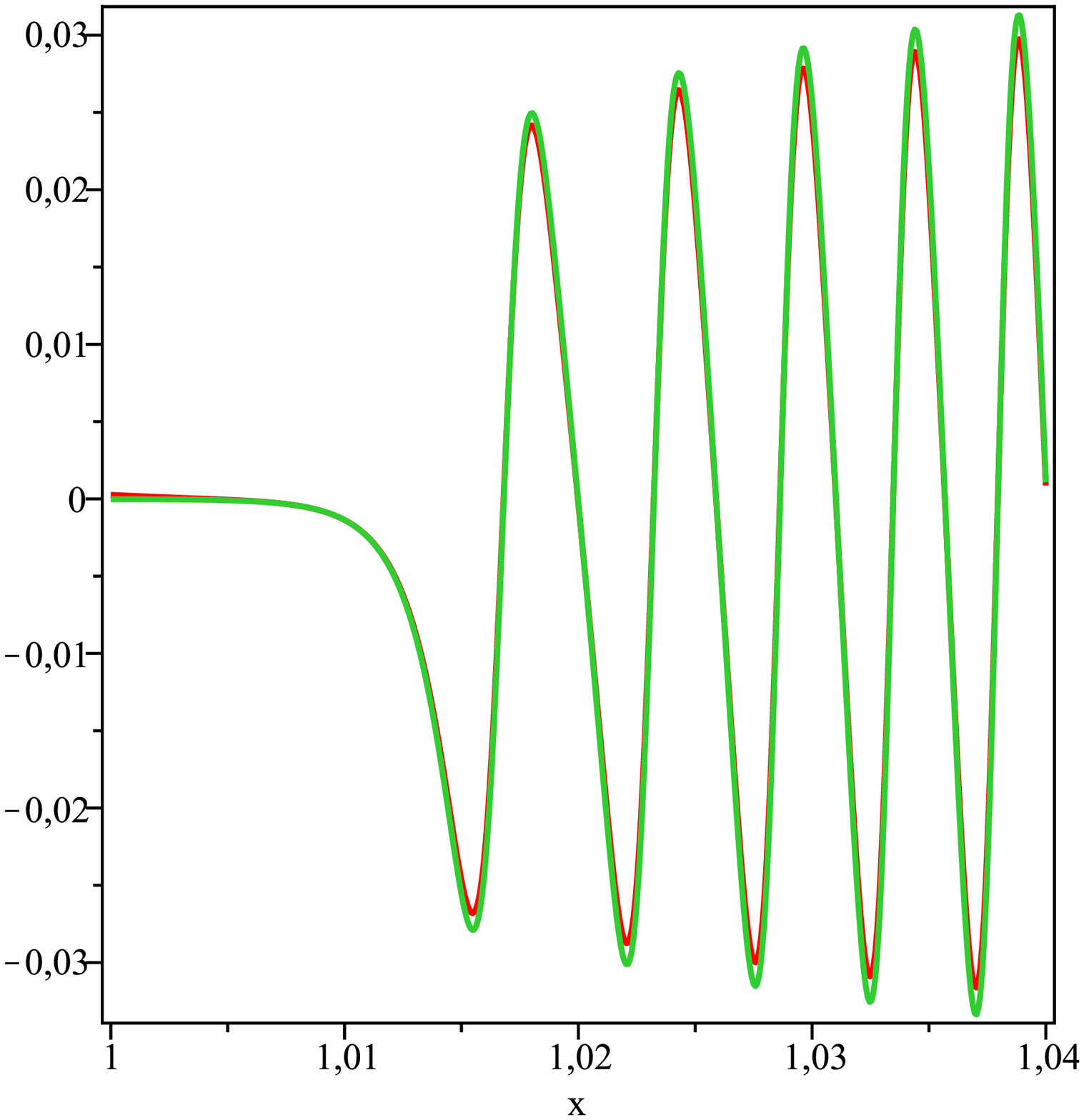}
\includegraphics[width=7.1cm]{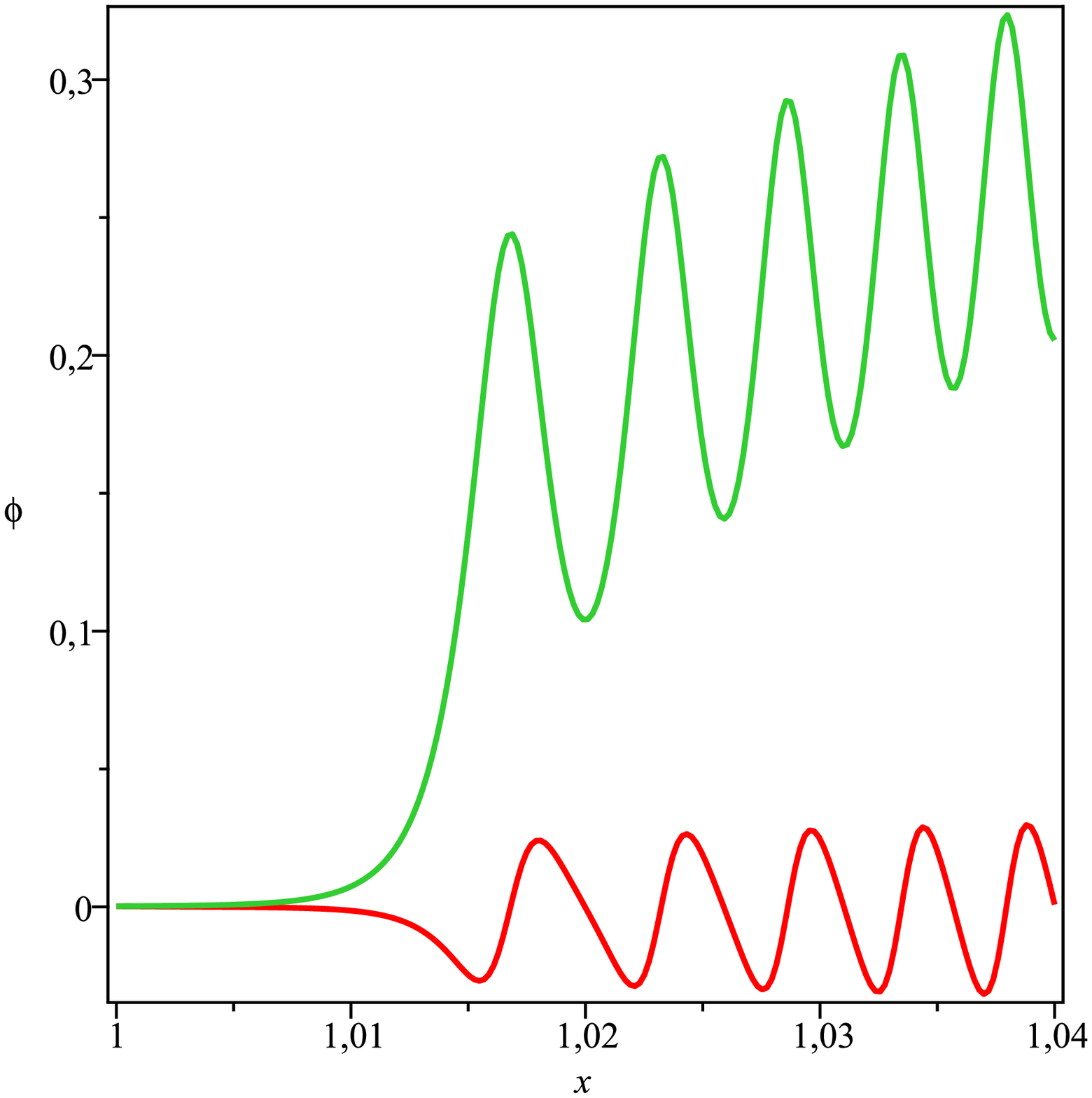}
\includegraphics[width=7.1cm]{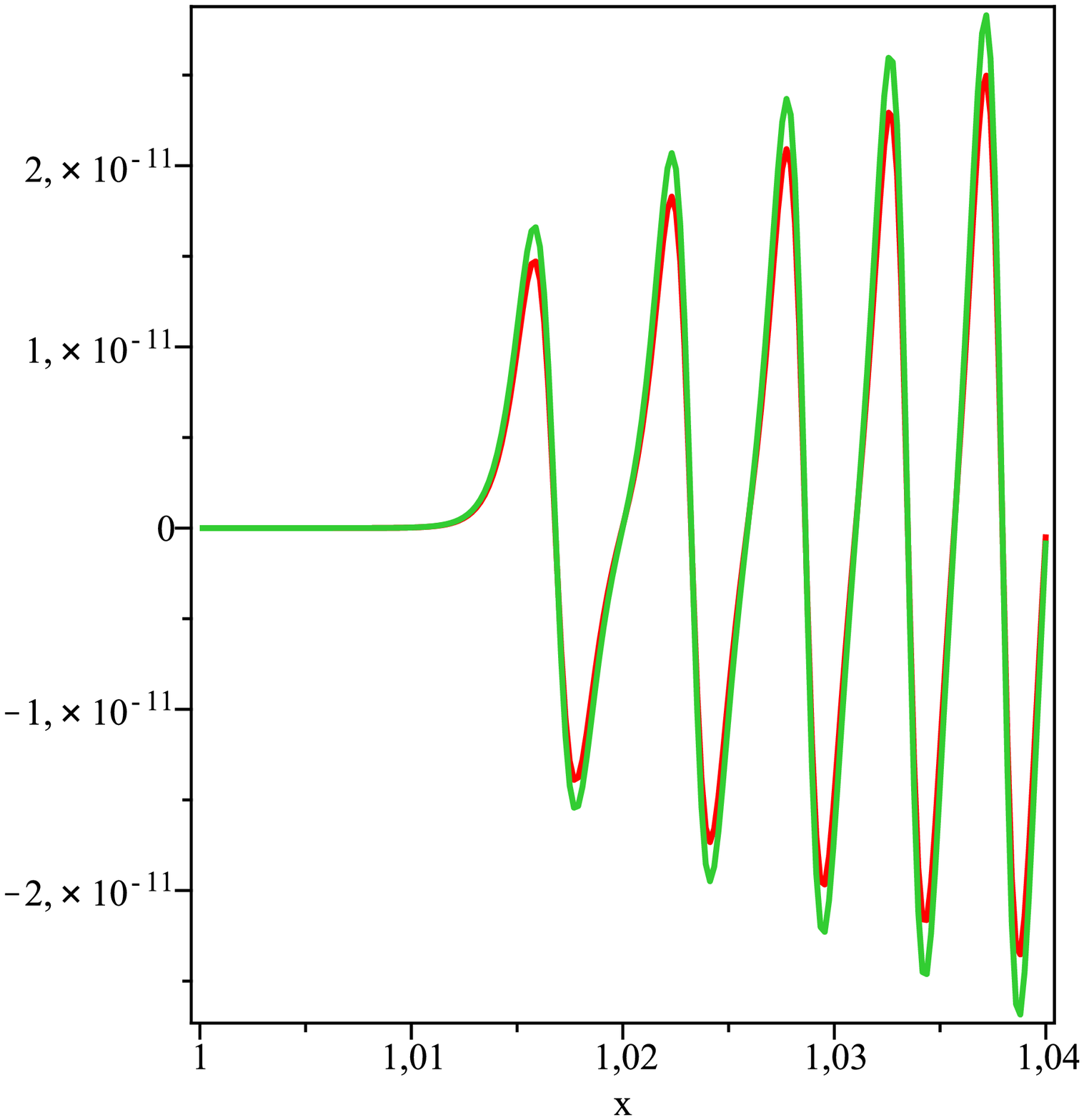}
\includegraphics[width=7.1cm]{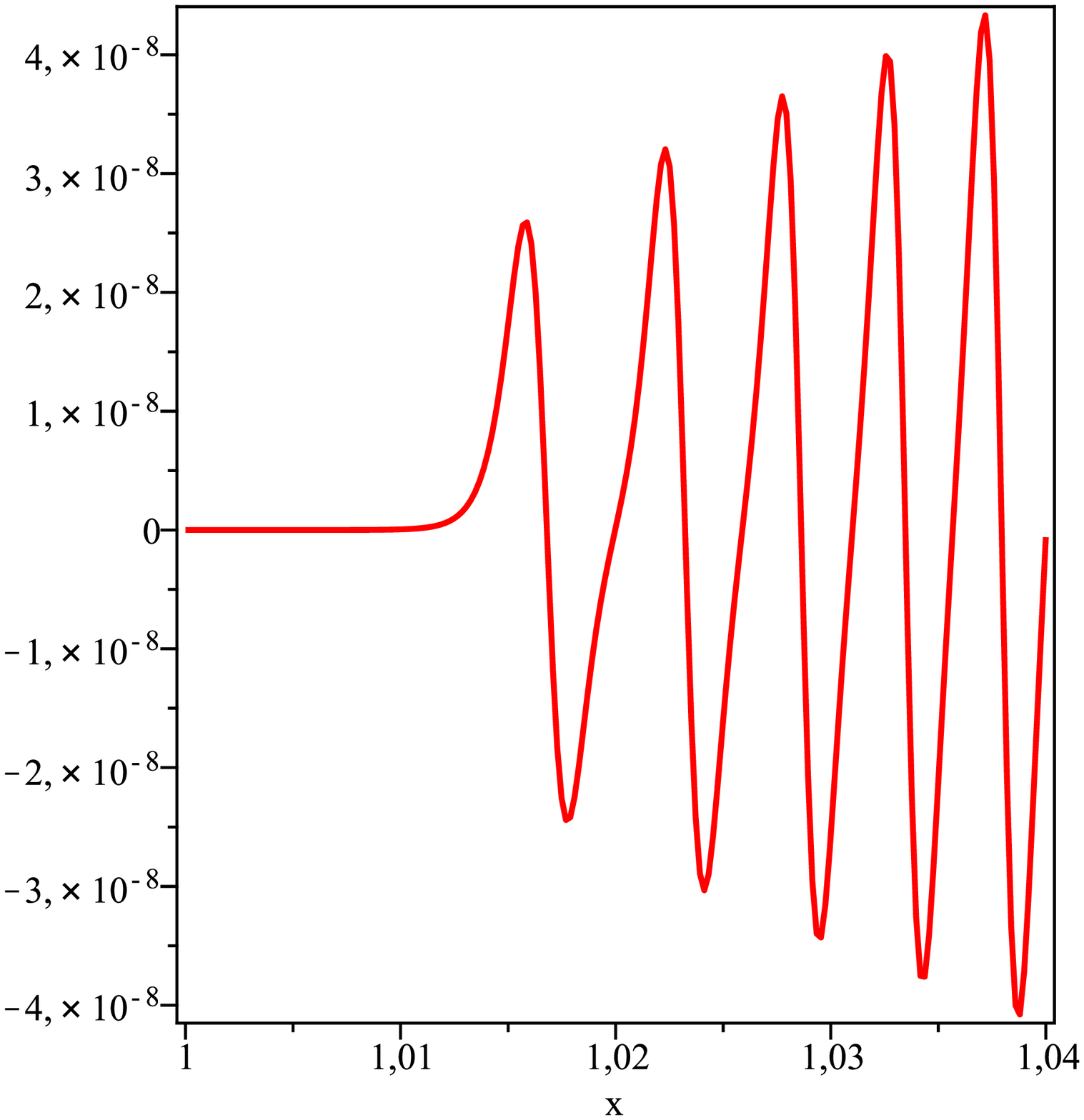}
\caption[]{The numerical results and analytic approximations for the perturbation evolution in the symmetron model, for which the model parameters are chosen as $\kappa=4\times10^3$, $z_\star\equiv z(t_\star)=3$ and $\alpha_\star =1$.  The initial condition is chosen as $\varphi(z_{eq})=10^{-2}$. {\it Upper left panel}: analytic (green curve) and numerical (red curve) solutions to the symmetron perturbation $\delta\varphi\equiv\delta\phi/\phi_\star$; note the excellent agreement between the two. {\it Upper right panel}: the time evolution of $\varphi$ (green curve) compared to that of $\delta\varphi$; note that $|\delta\varphi|$ is always smaller than $\varphi$, and does not blow up exponentially. {\it Lower left panel}: the analytic (green curve) and numerical (red curve) solutions to the $\Delta$ defined in the text; note the excellent agreement between the two. {\it Lower right panel}: the ratio $\Delta/\delta$ which is much less than unity, showing that the correction to the growth of matter perturbation due to the tachyonic instability is negligible.
\label{fig:techyonic}}
\end{center}
\end{figure*}

Defining $\tilde{\delta}\equiv x^{2\over3}\delta$, the growth equation becomes
\begin{eqnarray}
\tilde{\delta}''-\frac{4}{9x^2}\tilde{\delta} &\approx& -\frac{2\omega^2}{\xi_k^2}(\varphi\delta\varphi''+2\varphi'\delta\varphi'+\varphi''\delta\varphi),
\end{eqnarray}
the solutions of which are 
\begin{eqnarray}
\tilde{\delta} &=& \alpha_1 x^{4\over3}+\frac{\beta_1}{x^{1\over3}}+\frac{6\omega^2}{5\xi_k^2}\int_{x_k}^x\left(\frac{s^{4\over3}}{x^{1\over3}}-\frac{x^{4\over3}}{s^{1\over3}}\right)(\varphi\delta\varphi)''(s)ds,\nonumber
\end{eqnarray}
where $\alpha_1$ and $\beta_1$ are integration constants, and $x_k$ is the value of $x$ when the $k$-mode under consideration enters the horizon. As a result,
\begin{eqnarray}
\delta(x) &=& \alpha_1 x^{2\over3}+\frac{\beta_1}{x}+\frac{6\omega^2}{5\xi_k^2}\int_{x_k}^x\left(\frac{s^{4\over3}}{x}-\frac{x^{2\over3}}{s^{1\over3}}\right)(\varphi\delta\varphi)''(s)ds.\nonumber
\end{eqnarray}
Integrating by parts
\begin{eqnarray}
&&\int_{x_k}^x\left(\frac{s^{4\over3}}{x}-\frac{x^{2\over3}}{s^{1\over3}}\right)(\varphi\delta\varphi)''(s)ds\nonumber\\
&=& \left[\left(\frac{s^{4\over3}}{x}-\frac{x^{2\over3}}{s^{1\over3}}\right)(\varphi\delta\varphi)'(s)\right]_{x_k}^x\nonumber\\
&&-\int_{x_k}^x\left(\frac{4}{3}\frac{s^{1\over3}}{x}+\frac{1}{3}\frac{x^{2\over3}}{s^{4\over3}}\right)(\varphi\delta\varphi)'(s)ds.
\end{eqnarray}
At the horizon crossing the symmetron perturbations are taken to vanish, as during inflation the mass of the symmetron is much larger than the Hubble expansion rate. This implies that the initial conditions for the symmetron perturbation are
\begin{eqnarray}
\delta\phi_{t_k}\ \approx\ 0,\ \dot\delta\phi_{t_k}\ \approx\ 0
\end{eqnarray}
where $t_k$ is the horizon-entry time for the mode $k$
\begin{eqnarray}
t_k &=& \frac{8}{27 k^3 t_\star^2}
\end{eqnarray}
Using the fact that the tachyonic growth starts at $x=1$, we have
\begin{eqnarray}
&&-\int_{x_k}^x\left[\frac{4}{3}\frac{s^{1\over3}}{x}+\frac{1}{3}\frac{x^{2\over3}}{s^{4\over3}}\right](\varphi\delta\varphi)'(s)ds\nonumber\\
&\approx& -\int_{1}^x\left[\frac{4}{3}\frac{s^{1\over3}}{x}+\frac{1}{3}\frac{x^{2\over3}}{s^{4\over3}}\right](\varphi\delta\varphi)'(s)ds
\end{eqnarray}
The variation of $\varphi\delta\varphi$ is much faster than the other terms in this integral, so that the terms in the brackets could be absorbed into the derivative with respect to $x$, and the integrand becomes a total derivative. This implies that the growth factor behaves like
\begin{eqnarray}
\delta(x) &\approx& \left\lbrace\begin{array}{l}
\delta_k\left(\frac{x}{x_k}\right)^{2\over3}~~{\rm if}~x<1 \\
\delta_k\left(\frac{x}{x_k}\right)^{2\over3}-\frac{2\omega^2}{\xi_k^2x^{2\over3}}\varphi\delta\varphi(x)~~{\rm if}~x\geq1.
\end{array} \right.
\end{eqnarray}
As structures grow anomalously due to the tachyonic instability, we define the deviation from the normal growth as
\begin{eqnarray}
\Delta(x) &\equiv& \delta(x)-\delta_k\left(\frac{x}{x_k}\right)^{2\over3}.
\end{eqnarray}
When $x>1$, this becomes (remember that we have normalised $a_\star=1$)
\begin{eqnarray}
\Delta\ =\ -\frac{2\omega^2}{\xi_k^2x^{2\over3}}\varphi\delta\varphi(x)\ =\ -\frac{9}{2}\frac{a^2H^2}{k^2}\alpha_{\phi}\frac{\delta\phi}{m_{pl}}.
\end{eqnarray}
Our numerical and analytic solutions to $\Delta$ are shown in the lower left panel of Fig.~\ref{fig:techyonic}, and the agreement is very good. To see how big an impact the tachyonic instability could have on the growth of matter density perturbation, let us consider the quantity
\begin{eqnarray}
\frac{\Delta}{\delta_{\rm GR}}\ =\ \frac{9}{2}\frac{a^2H^2}{k^2\delta_{\rm GR}}\alpha_{\phi}\frac{\delta\phi}{m_{pl}}\ =\ -\frac{27}{4}\left(\frac{aH}{k}\right)^4\frac{\alpha_\phi\frac{\delta\phi}{m_{\rm pl}}}{\Phi_{N,{\rm GR}}}\ \ \
\end{eqnarray}
where in the second step we have used the Poisson equation Eq.~(\ref{poisson}) in GR, i.e., with the $\delta\phi$ term removed. To have a rough idea about the magnitude of this quantity, we need the following ingredients:
\begin{enumerate}
\item During the brief period of tachyonic instability, not only $\delta\phi$ but also the symmetron field $\phi$ itself grows exponentially. In particular, the mass-squared for $\phi$, $m^2(\phi)$, is more negative than  for $\delta\phi$ because the latter is $k^2/a^2+m^2(\phi)\geq m^2(\phi)$ for all $k$-modes. As a result, although the growth of $\delta\phi$ is unstable, it is still slower than that of $\phi$ and we expect $\delta\phi/\phi < \mathcal{O}(1)$, as can been seen from the upper right panel of Fig.~\ref{fig:techyonic}.
\item We have seen above that for the symmetron model to evade the local experimental constraints we must have $\phi/M<\phi_\star/M\sim\mathcal{O}(10^{-3})$, as well as $M/m_{\rm pl}\leq\mathcal{O}(10^{-3})$. As a result, $\phi/m_{\rm pl}\leq\mathcal{O}(10^{-6})$.
\item $k/aH\gg1$ for the sub-horizon modes and $k/aH\sim\mathcal{O}(1)$ for near-horizon modes.
\item $\Phi_{N,{\rm GR}}$ is the gravitational potential in general relativity, which should be much less than unity, and we could use $\Phi_N\sim\mathcal{O}(10^{-7} - 10^{-3})$ for conservative estimates.
\item $\alpha_\phi\sim\mathcal{O}(1)$.
\end{enumerate}
For sub-horizon modes, taking $k/aH\sim50$ and $\Phi_{N,{\rm GR}}\sim10^{-4}$, we find that $|\Delta/\delta_{\rm GR}|\sim\mathcal{O}(10^{-8})$ (the exact value, of course, depends on the values of $\phi, M$ and $\delta\phi$), which is consistent with the numerical result shown in the lower right panel of Fig.~\ref{fig:techyonic}. For near horizon modes, on the other hand, $k/aH\sim\mathcal{O}(1)$ and thus we have $|\Delta/\delta_{\rm GR}|\sim\mathcal{O}(10^{-6}/\Phi_{N,{\rm GR}})$, i.e., the very large scale growth of matter perturbation could be modified significantly and leave potentially observable signatures\footnote{Note that even on horizon scales where the tachyonic instability is supposed to be more severe, we will not have an unbounded blow-up of the matter perturbation as naively suggested by the equation of motion for the symmetron perturbation. Note also that near the horizon the approximations used to derive the results in this subsection might be inaccurate, but we expect them to provide a reasonable order of magnitude estimate.} if $|\Phi_{N,{\rm GR}}|\lesssim10^{-6}$. In the present paper, however, we will be primarily interested in the sub-horizon scales and neglecting the tachyonic instability is a reasonably good approximation.

In fact, a very accurate approximation for $\delta\varphi$ can be obtained by noticing that $\dot \phi$ satisfies
\begin{equation} \dot{\ddot{\phi}}+3H\ddot{\phi}+\left(-\frac{9}{2}H^2+m^2(\phi)\right)\dot{\phi}=0 \end{equation}
while  $\delta\phi$ is a solution of
\begin{equation} \ddot{\delta\phi}+3H\dot{\delta\phi}+\left(\frac{k^2}{a^2}+m^2(\phi)\right)\delta\phi=-2\Phi_N V'(\phi)+4\dot{\Phi}_N\dot{\phi}-\frac{\bar{\rho}\phi}{M^2}\delta \end{equation}
As $H^2$ is small compared to $m^2(\phi)$, $\delta\phi$ and  $\dot{\phi}$ satisfy the same equation provided  $\frac{k^2}{a^2}\ll m^2(\phi)$ and one can neglect the right hand side of the Klein-Gordon equation. As $m^2(\phi)=O(\mu^2\sim10^6H_0^2)$  and the tachyonic modes are  smaller  than $k\lesssim  60H_\star$, the first condition is satisfied. Moreover, the  right-hand side is negligible provided $\delta\ll\delta\varphi$. In this case $\delta\phi$ and  $\dot{\phi}$ satisfy the same equation provided  $\delta$ is small compared to $\frac{\delta\phi}{\phi_0}$. As a result we have\cite{Brax:2010ai}
\begin{equation}
\delta\varphi \approx \frac{\dot\phi}{\dot\phi_\star} \delta\varphi_\star
\end{equation}
This approximation is numerically very accurate for all $H_\star \lesssim k \lesssim 60 H_\star$ as can be seen in Fig.~4.

Why does the exponential blow-up of $\delta\phi$ have small effects on the growth of matter perturbation? First, the rapid growth of $\delta\phi$ is associated with an even faster growth of $\phi$ and so the symmetron perturbation does not really enter the nonlinear regime (see upper right panel of Fig.~\ref{fig:techyonic}). In the background cosmology, the blow-up of $\phi$ has not spoiled the matter dominated era, and certainly we should not expect an even slower perturbation growth to be too much problematic in the linear regime. Second, as mentioned many times above, the tachyonic period is very brief, thanks to the exponential growth of $\phi$ which quickly settles down into the non-tachyonic regime; as a result the blow-up of either $\phi$ or $\delta\phi$ simply cannot accumulate enough momentum to have big impacts on the structure formation.

The study of the structure formation of the symmetron model in nonlinear regime is beyond the scope of present paper and numerical simulations have been performed in \cite{wmld2011}. Be aware that an accurate study of the tachyonic behaviour of the model will involve a very big simulation box (to include the near horizon scales) and the full temporal evolution of $\delta\phi$, which is not achievable by known $N$-body simulations at present.

\section{Structure Formation in the Adiabatic Approximation}

\label{sect:linpert_numeric}

Most  of the sub-horizon scales are in the non-tachyonic regime  while  the symmetron field  oscillates around the (moving) minimum of the effective potential. This means that the adiabatic approximation, where the scalar field (in background cosmology) always follows the minimum, provides a good description of perturbation theory. This has been verified by numerical simulations \cite{wmld2011}. In this section we shall study the linear structure formation quantities in the symmetron model under the adiabatic approximation.

\subsection{The Adiabatic approximation}

In the adiabatic approximation, we neglect the effect of the short time period when the symmetron perturbation becomes tachyonic which, as we have seen, is justified. In this case we have
\begin{eqnarray}
\phi_{\rm min}(\rho) &=& \phi_{\star} \sqrt{1 - \frac{\rho}{\rho_{\star}}} \theta\left(\rho_{\star} -  \rho\right),  \\
\alpha_{\rm min}(\rho) &=& \alpha_{\star} \left(1-\frac{\rho}{\rho_{\star}}\right) \theta\left(\rho_{\star} - \rho\right),
\end{eqnarray}
where $\theta(x)$ is the Heaviside function, and
\begin{eqnarray}
m^2_{\rm min}(\rho) = \left\{
  \begin{array}{cc}
  \mu^2 \left[ \frac{\rho}{\rho_{\star}} - 1\right]  & \rho > \rho_{\star}, \\
  \\
  2\mu^2 \left[ 1-\frac{\rho}{\rho_{\star}}\right] &  \rho < \rho_{\star}
  \end{array}  \right.
\end{eqnarray}
The symmetron model is characterised by the dimensionless parameters $z_{\star}$,  $A_{2}$ and $\alpha_{\star}$ determining the redshift at the transition
\begin{eqnarray}
(1+z_{\star})^3 &=& \frac{\rho_{\star}}{\rho_{\rm m0}}
\end{eqnarray}
In terms of these parameters, $\mu^2 = 3 A_2 \Omega_{\rm m0} H_0^2 (1+z_{\star})^3$ where $\Omega_{\rm m0}H_0^2 = \kappa_4^2 \rho_{\rm m0}/3$. We already know that local tests of gravity require $A_{2} \gtrsim 10^{6}$.  For $z_{\star} > 0$ this implies that $\mu^2 \gtrsim 10^{6}H_0^2 \gg H_0^2$ and therefore the symmetron mass is always much larger than the Hubble rate.

We define $\bar{\phi}(t)$ to be the value of the symmetron field in the cosmological background.  The symmetron equation of motion and Einstein equation in the cosmological background reduce to
\begin{eqnarray}
-\ddot{\bar{\phi}} - 3H \dot{\bar{\phi}} &=& -\mu^2 \bar{\phi} + \lambda \bar{\phi}^3 + \frac{\bar{\phi}}{M^2} \rho_{\rm m}(a), \\
H^2 &=& \frac{\kappa_4^2 }{3}\left[\bar{\rho}_{\rm m}(a) + \bar{\rho}_{\rm rad}(a)  + V_0 + \bar{\rho}_{\phi} \right], \\
\bar{\rho}_{\phi} &=& \frac{\bar{\phi}^2}{2M^2}\rho_{\rm m}(a) -  \frac{1}{2}\mu^2 \bar{\phi}^2 + \frac{1}{4}\lambda \bar{\phi}^4 + \frac{1}{2} \dot{\bar{\phi}}^2.\ \ \
\end{eqnarray}
where $\rho_{\rm m}(a) \propto a^{-3}$ and $\rho_{\rm rad}(a) \propto a^{-4}$.  Because $\mu^2 \gg H^2$ today, one finds that in the cosmological background $\bar{\rho}_{\phi} \ll \bar{\rho}_{\rm m}$ and in the adiabatic approximation the cosmological symmetron $\bar{\phi}$ tracks the minimum of the effective potential, which is a cosmological attractor:
\begin{eqnarray}
\bar{\phi}\ \approx\ \phi_{\rm min}(\bar{\rho}_{\rm m})\ =\ \phi_{\star} \sqrt{1 - \frac{\bar{\rho}_{\rm m}}{\rho_{\star}}} \theta\left(\rho_{\star} -  \bar{\rho}_{\rm m}\right).
\end{eqnarray}
In terms of the redshift $z$, this becomes
\begin{eqnarray}
\bar{\phi} &\approx& \phi_{\star} \sqrt{1 - \left(\frac{1+z}{1+z_{\star}}\right)^3} \theta(z_{\star}-z),
\end{eqnarray}
Since $\bar{\rho}_{\phi}$ is negligible, the background cosmological behaviour of the symmetron model  is just the unmodified $\Lambda$CDM model with a cosmological constant $\Lambda = \kappa_4^2 V_0$:
\begin{eqnarray}
H^2 &\approx& \frac{\kappa}{3} \left[ \bar{\rho}_{\rm rad}(a) + \bar{\rho}_{\rm m}(a)\right] + \frac{\Lambda}{3}. \nonumber
\end{eqnarray}
We define $m(\phi) = \bar{m}(z)$ and $\alpha(\phi) = \bar{\alpha}(z)$ in the cosmological background and then we have:
\begin{eqnarray}
\bar{m}^2(z) &\approx&   \left\{
  \begin{array}{cc}
  3A_{2} \Omega_{\rm m0} H^2_0\left[(1+z)^3-(1+z_{\star})^3\right]  & z > z_{\star}, \\
  \\
  6A_{2} \Omega_{\rm m0} H^2_0 \left[(1+z_{\star})^3-(1+z)^3\right] &  z < z_{\star}.
  \end{array}  \right. , \\
\bar{\alpha}(z) &\approx& \left\{
\begin{array}{cc}
  0  & z > z_{\star}, \\
  \\
  \alpha_{\star}\left[1 - \left(\frac{1+z}{1+z_{\star}}\right)^3\right]  &  z < z_{\star}.
  \end{array}  \right.
\end{eqnarray}
This allows one to study the growth of structure in the adiabatic approximation.

\subsection{Linear Structure Formation}

We assume that high density localised clumps of matter, e.g. galaxies, have $\rho \gg \rho_{\star}$ and therefore are effectively screened from the symmetron force.  We define $\delta_{\rm g}(\mathbf{x},t)$ as the relative large scale coarse grained density perturbation in the galaxies, and $\delta_{\rm m}(\mathbf{x},t)$ to be the large scale (linear) density perturbation in the smooth cold dark matter. We work to linear order in $\delta_{\rm g} \ll 1$ and $\delta_{\rm m} \ll 1$. We define $\tilde{\delta}_{\rm g}(\mathbf{k},t)$ and $\tilde{\delta}_{\rm m}(\mathbf{k},t)$ to be the Fourier transforms of the relative density perturbations.

Using the adiabatic approximation for the background evolution of the symmetron field, and taking $p = -\ln (1+z)$, we have (see e.g. \cite{Brax:2009ab}):
\begin{eqnarray}
\tilde{\delta}_{{\rm g},pp} + \left[2-\frac{3\Omega_m}{2}\right]\tilde{\delta}_{{\rm g},p} &=& \frac{3}{2}\Omega_m \tilde{\delta}_{\rm m}, \\
\tilde{\delta}_{{\rm m},pp} + \left[2-\frac{3\Omega_m}{2}\right]\tilde{\delta}_{{\rm m},p} &=& \frac{3}{2}\left[1+ \frac{\bar{\alpha}(z)}{1+ (1+z)^{-1} k^{-2}\bar{m}^2(z)} \right]\Omega_m \tilde{\delta}_{\rm m},\ \nonumber
\end{eqnarray}
where $_{,p}\equiv\partial/\partial p$, and
\begin{eqnarray}
\Omega_{\rm m}\ =\ \frac{\kappa_4^2 \rho_{\rm m}(a)}{H^2}\ =\ \frac{\Omega_{\rm m0}(1+z)^3 H_0^2}{H^2}. \nonumber
\end{eqnarray}

Galaxies are often used as observational tracers of the linear dark matter perturbation.  The latter feels both gravity and the fifth force due to the symmetron coupling whilst the former only evolves under gravity. This is very similar to the chameleon and environmentally dependent dilaton scenarios \cite{kw2004a,kw2004b,ms2006,ms2007,lb2007,bbds2008}. Working in the Fourier space, we have solved the relevant perturbation equations numerically to find the relevant quantities (for simplicity, in what follows we shall neglect the tilde of $\tilde{\delta}_{\rm g}$ and $\tilde{\delta}_{\rm m}$)
\begin{itemize}
\item The growth rate $f_{\rm gal} =  d(\ln \delta_{\rm g}) /d\ln a$.
\item The slip function, $\Sigma_{\kappa m}$, measured by weak gravitational lensing.
\item The slip function $\Sigma_{\kappa I}$ measured by the integrated Sachs-Wolfe (ISW) effect.
\item The indicator of modified gravity $E_G$. 
\end{itemize}
Let us first recall their definitions.
In a modified gravity setting, there are  two metric potentials $\Phi$ and $\Psi$:
\begin{eqnarray}
ds^2 &=& a^2(\eta)\left[-(1+2\Psi)d\eta^2+ (1-2\Phi)dx^2\right]\nonumber
\end{eqnarray}
In the symmetron model here, the absence of anisotropic stress implies
\begin{eqnarray}
\Phi_N\ =\ \Phi\ =\ \Psi\nonumber
\end{eqnarray}
as in Eq.~(\ref{eq:metric}).
Weak lensing is sensitive to  $\Phi+\Psi$ while the ISW effect is proportional to $\dot{\Phi}+\dot{\Psi}$.  The slip functions are defined by
\begin{eqnarray}
k^2(\Phi+\Psi) &=& -8\pi G a^2 \bar{\rho} \Sigma_{\kappa m} D_{\rm GR}\delta_{\rm i}, \\
H^{-1} k^2 (\dot{\Phi}+\dot{\Psi})&=& -8\pi G a^2 \bar{\rho} \Sigma_{\kappa I} (f_{\rm GR}-1) D_{\rm GR}\delta_{\rm i},
\end{eqnarray}
in which $\delta_{i}$ is the primordial matter density perturbation (measured from the CMB), $D_{\rm GR}$ is the growth factor in GR, $\delta_{\rm m} \equiv D_{\rm GR}\delta_{i}$ is the GR density contrast and $f_{\rm GR} \equiv d\ln D_{\rm GR}/d\ln a$ is the GR growth rate.  In the symmetron model
$\delta_{\rm m} \equiv D_{\rm m} \delta_{i}$ and so
\begin{eqnarray}
\Sigma_{\kappa m} &=& D_{\rm m}/D_{\rm GR}
\end{eqnarray}
and
\begin{eqnarray}
\Sigma_{\kappa I} &=& (f_{\rm m}-1)\Sigma_{\kappa m}/(f_{\rm GR}-1)
\end{eqnarray}
where we have defined
\begin{eqnarray}
f_{\rm m} \equiv d\ln D_{\rm m} /d\ln a\ =\ d\ln \delta_{\rm m} /d \ln a.
\end{eqnarray}
We also define the linear, bias corrected, growth rate for the galaxies. This is defined by assuming that
\begin{eqnarray}\delta_{\rm g} &=& D_{\rm gal}^{bc} \delta_{i} + \Delta_{0}
\end{eqnarray}
where $\delta_{i}$ is the initial Gaussian perturbation, $D_{\rm gal}^{bc}$ is the bias corrected growth factor for galaxies, and the zero mode $\Delta_{0}$ is the source of the bias. We then define the bias corrected galaxy density contrast as
\begin{eqnarray}
\delta_{\rm g}^{bc} &=& D_{\rm gal}^{bc}\delta_{i}\ =\ b_{\rm lin}^{-1}(\delta_{g})\delta_{g}
\end{eqnarray}
where the linear bias is defined by
\begin{eqnarray}
b_{\rm lin}^{-1} = 1-\Delta_{0}/\delta_{g}.
\end{eqnarray}
$\Delta_{0}$ and hence the linear bias $b_{\rm lin}$ can be estimated directly from galaxy surveys using higher order statistics. We also define the bias corrected galaxy growth rate as
\begin{eqnarray}
f_{\rm gal}^{\rm bc} &\equiv& \frac{d\ln \delta_{\rm g}^{\rm bc}}{d\ln a}
\end{eqnarray}
In the absence of any deviations from GR, $f_{\rm gal}^{bc} = f_{\rm GR} \approx \Omega_{\rm m}^{0.545}$, and $\Sigma_{\kappa m} =\Sigma_{\kappa I} = 1$.
However in the symmetron model, the symmetron fifth force could result in deviations from these GR results.

 Finally, we consider the modified gravity sensitivity parameter defined by
\begin{eqnarray}
E_{\rm G} &=& \frac{k^2 (\Psi+\Phi)}{-3H_0^2 a^{-1}\theta_{\rm gal}} , \nonumber
\end{eqnarray}
where $\theta_{\rm gal} = -\dot{\delta}_{\rm gal}/H = -d\delta_{\rm gal}/d\ln a$, and so $\theta_{\rm gal} = -f_{\rm gal}^{\rm bc}D_{\rm gal}^{bc}\delta_{i}$ and with $a_0=1$ today:
\begin{eqnarray}
E_{\rm G} = \frac{\Omega_{\rm m0} D_{\rm m}}{f_{\rm gal}^{\rm bc}D_{\rm gal}^{bc}}
\end{eqnarray}
In GR, $E_{\rm G}^{(\rm GR)} = \Omega_{\rm m0}/f_{\rm GR} \approx \Omega_{\rm m0} \Omega_{\rm m}^{-0.545}$.

{The quantities $f_{\rm gal}$, $\Sigma_{\kappa m}$, $\Sigma_{\kappa I}$ and $E_{\rm G}$ could give us a feeling about how much the symmetron predictions deviate from the predictions
made in General Relativity.} We find that these quantities depend on the redshift, $z$, the dimensionless spatial scale parameter $q \equiv \vert\mathbf{k}\vert/A_2^{1/2} \Omega_{\rm m0}^{1/2} H_0$ and the model parameters $\alpha_{\star}$ and $z_{\star}$.  Recall that $z_{\star}$ is the redshift at which the symmetron mediated fifth force turns on, and $\alpha_{\star}$ is the relative strength of this extra force to that of gravity at late times (i.e., $(1+z) \ll (1+z_{\star})$) and on large scales.

\begin{figure*}[tbh]
\begin{center}
\includegraphics[width=7.5cm]{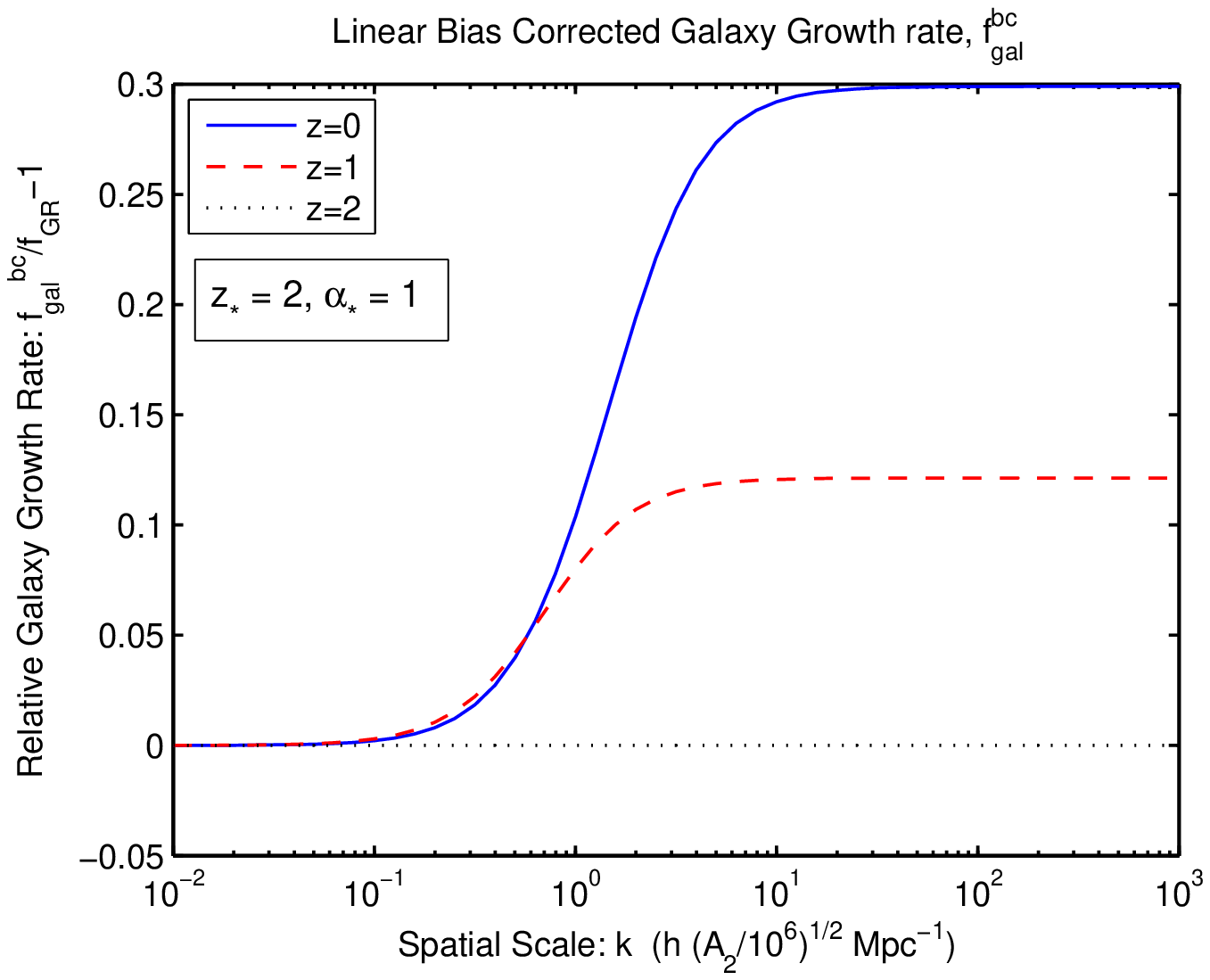}
\includegraphics[width=7.5cm]{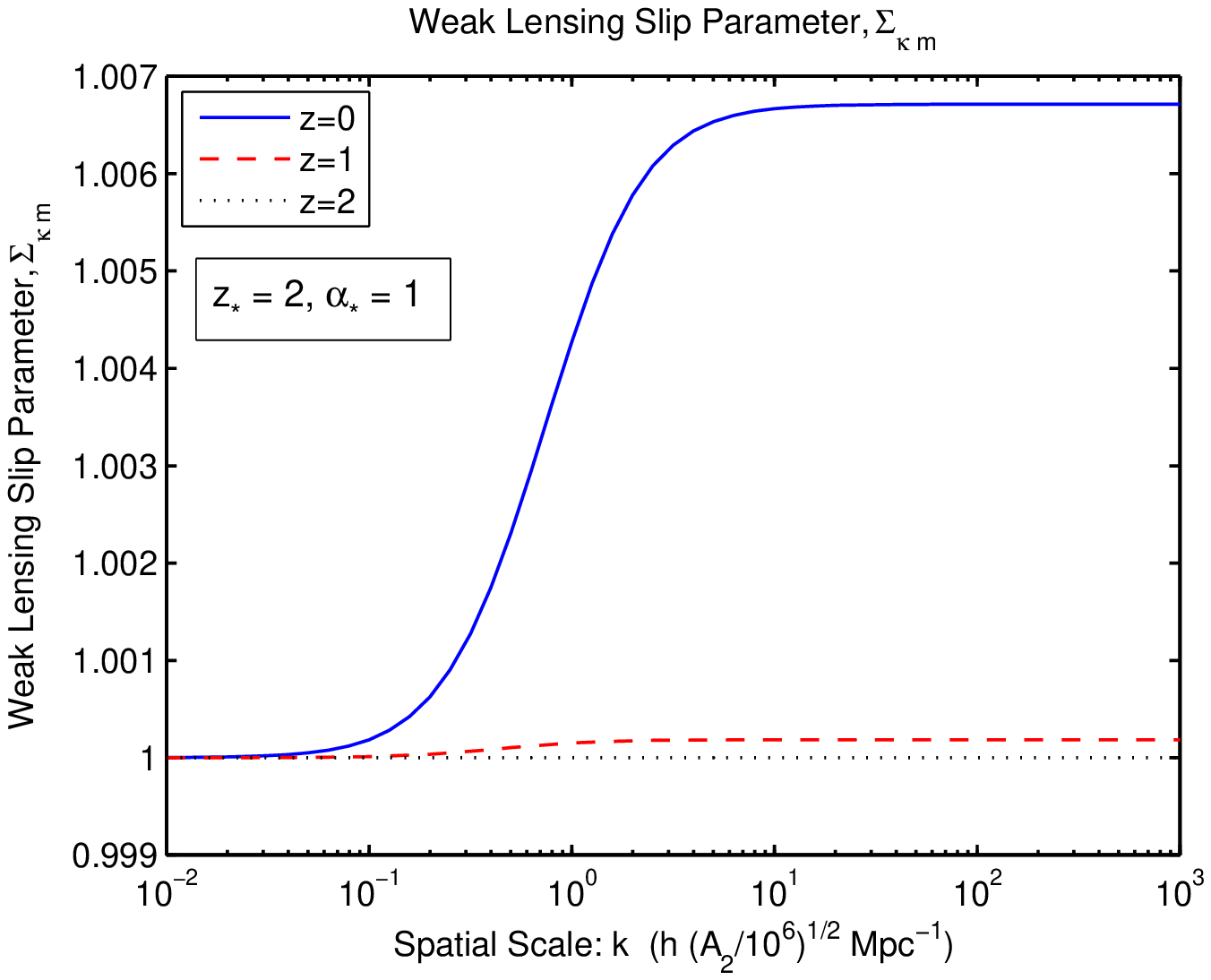}
\includegraphics[width=7.5cm]{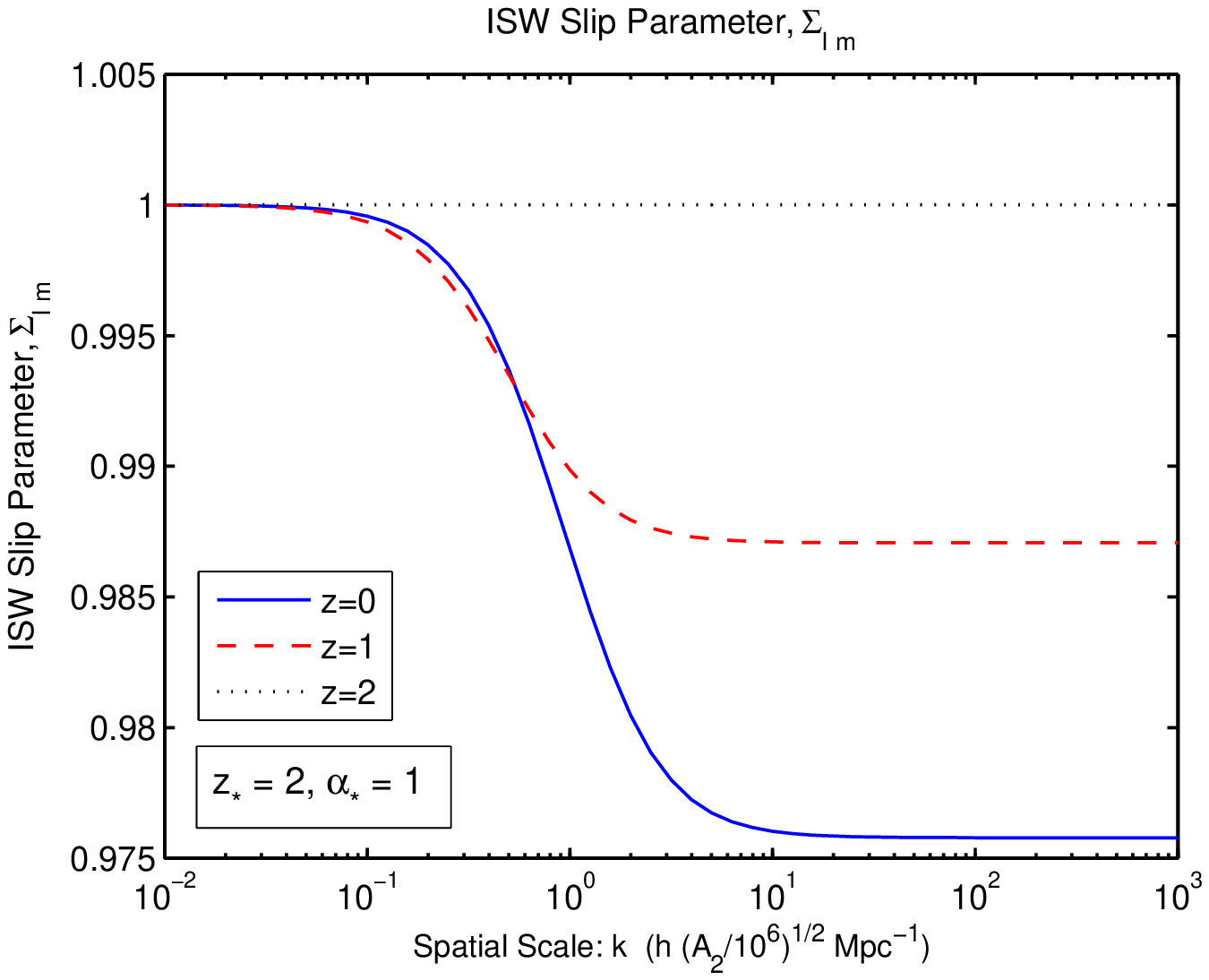}
\includegraphics[width=7.5cm]{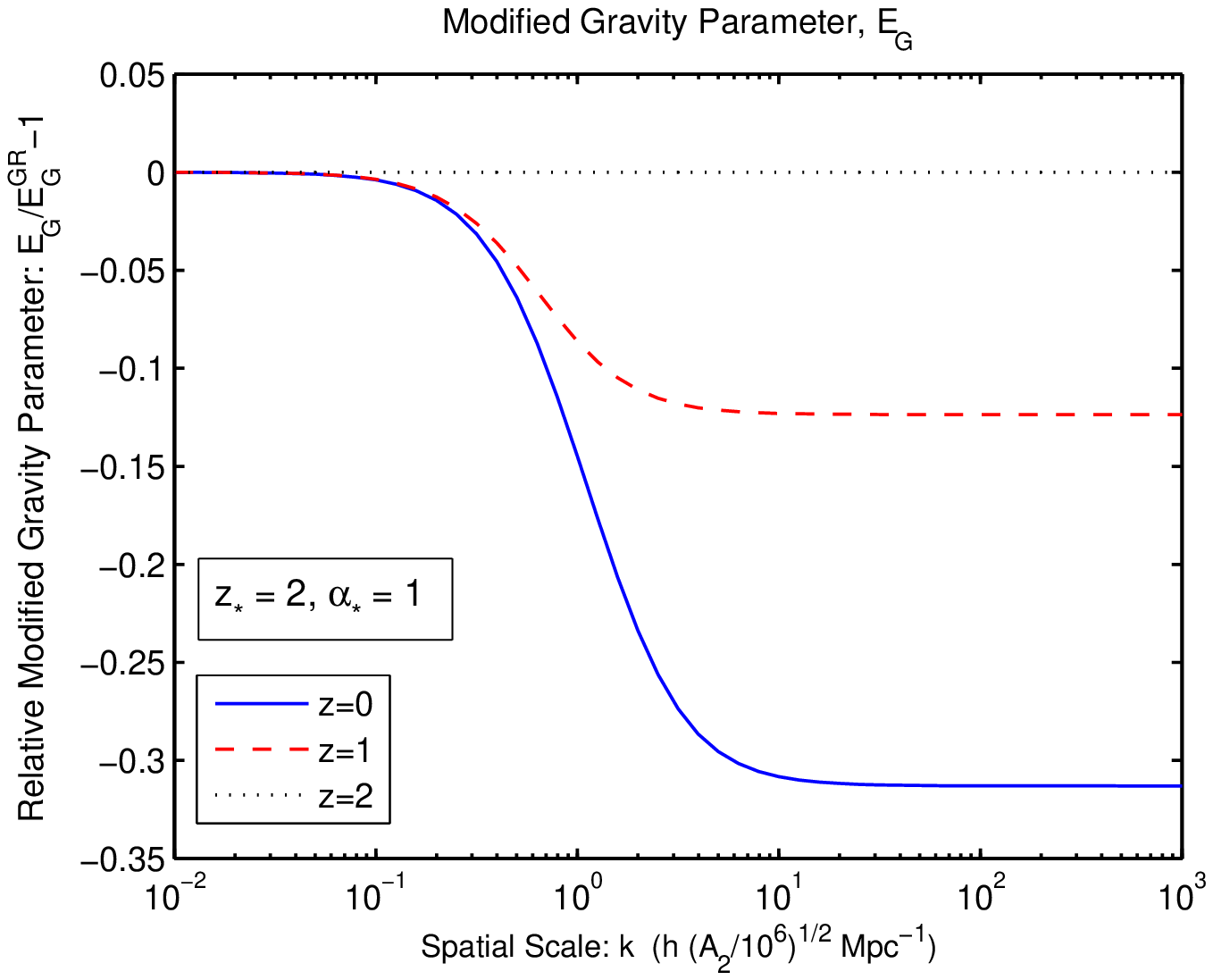}
\caption[]{Effect of the symmetron field on large scale structure formation.  From top to bottom the plots show: the predicted deviation of the (linear bias corrected) galaxy growth, $f_{\rm gal}^{\rm bc}$, from its GR value, $f_{\rm GR}$, the predicted value of the slip parameter, $\Sigma_{\rm \kappa m}$, extrapolated from weak lensing measurements, the predicted slip parameter extrapolated from ISW measurements, $\Sigma_{\rm I m}$, and the relative deviation of the modified gravity parameter, $E_{\rm G}$, from its GR value. These plots are for $[\alpha_{\star} = 1,\,\, z_{\star} = 2]$, and show the values of $f_{\rm gal}^{\rm bc}$, $\Sigma_{\rm \kappa m}$ and $\Sigma_{\rm Im}$ at the present day, $z=0$ (solid blue line), $z=1$ (dashed red line) and $z=2$ (dotted black line) for different values of the inverse spatial scale, $k$.
\label{fig1}}
\end{center}
\end{figure*}

\subsection{Numerical results}
\begin{figure*}[tbh]
\begin{center}
\includegraphics[width=7.1cm]{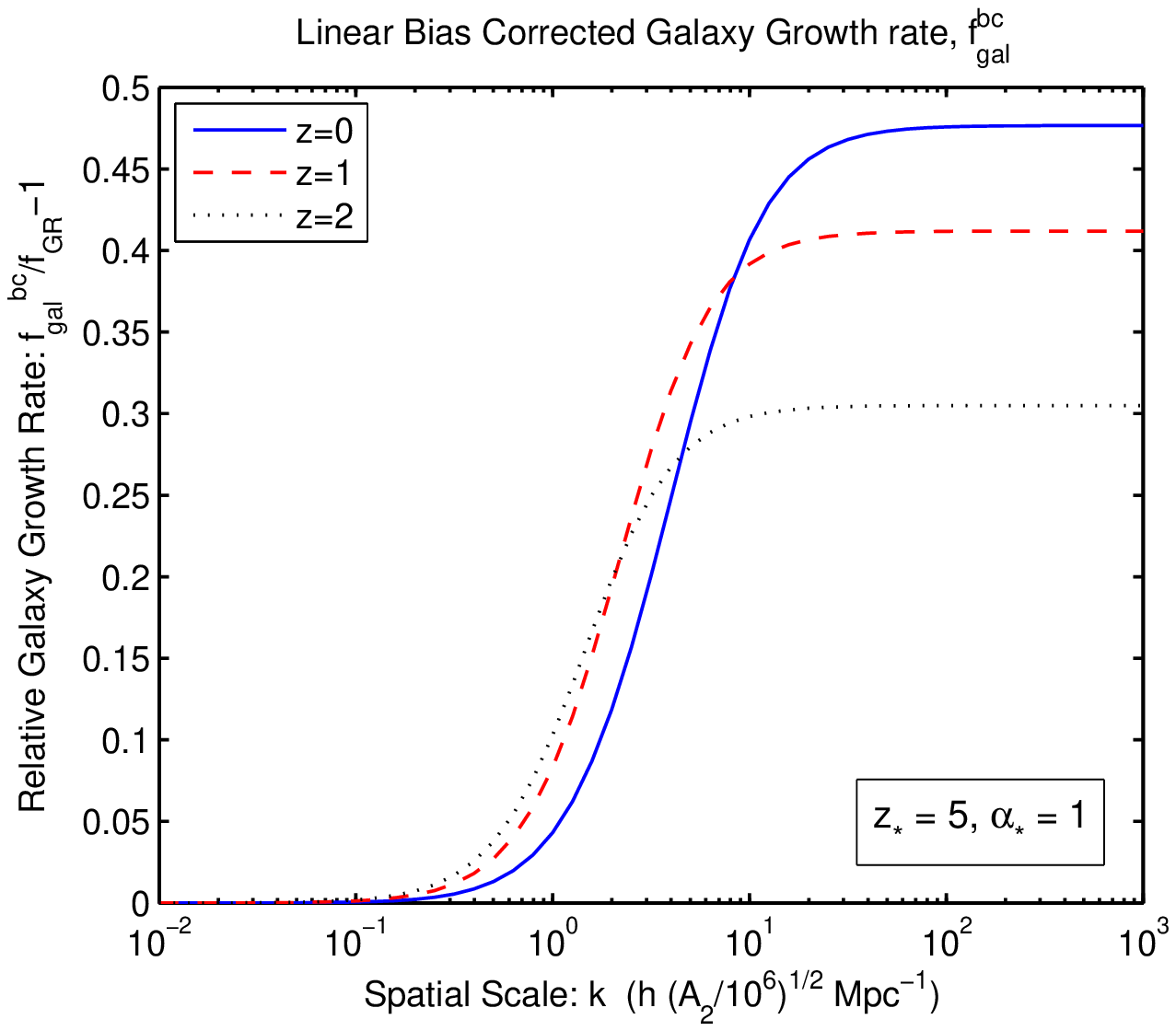}
\includegraphics[width=7.5cm]{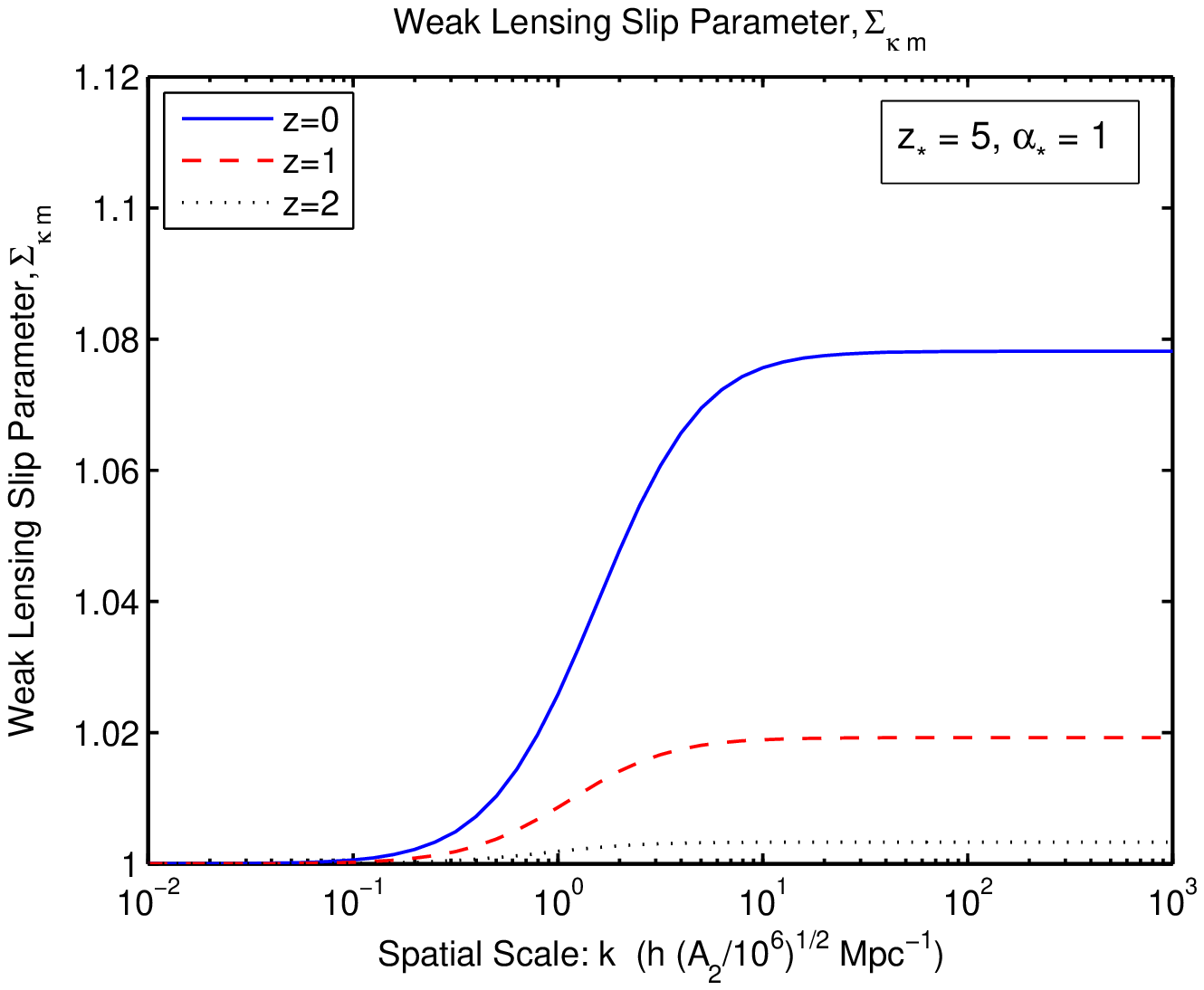}
\includegraphics[width=7.5cm]{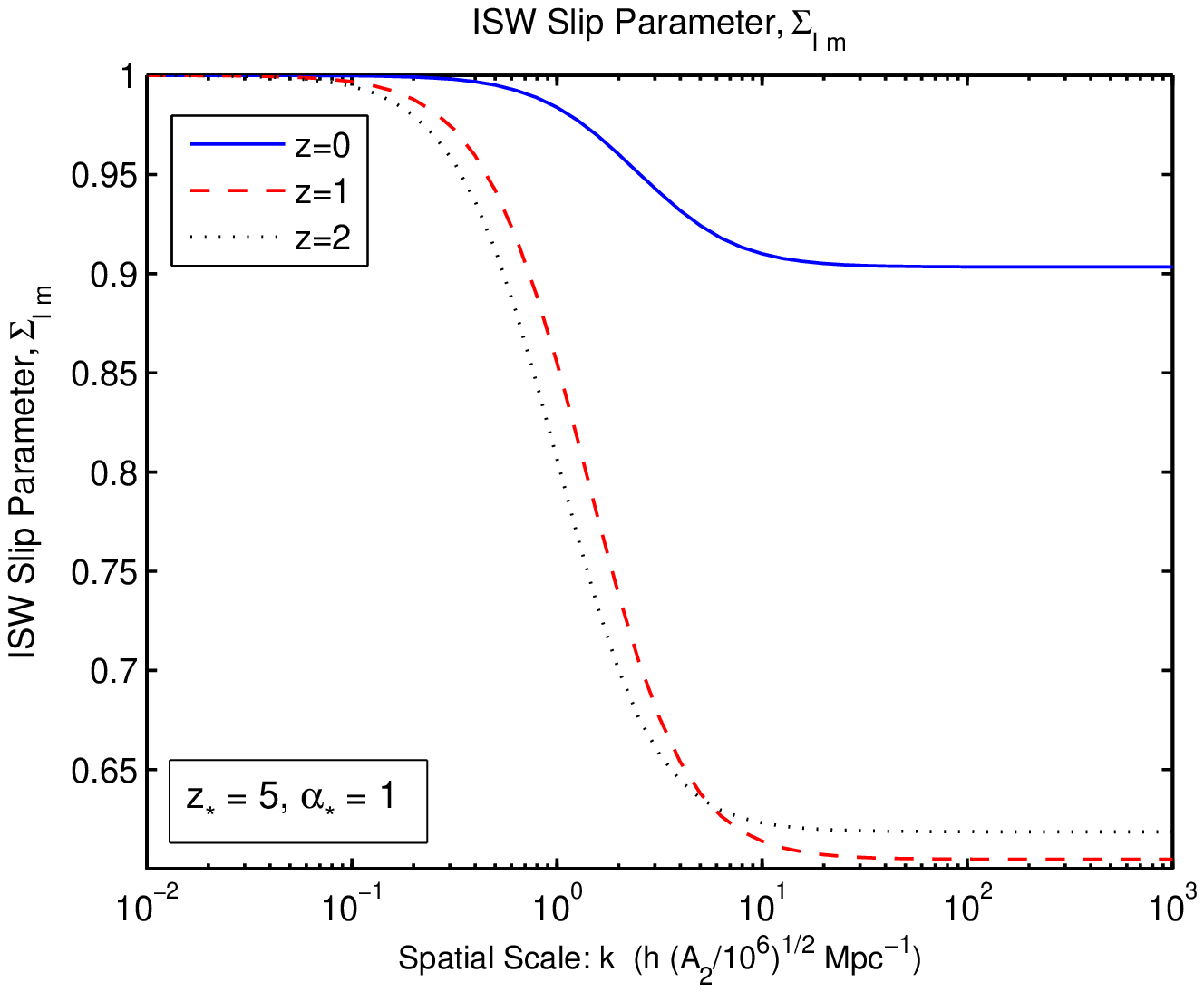}
\includegraphics[width=7.5cm]{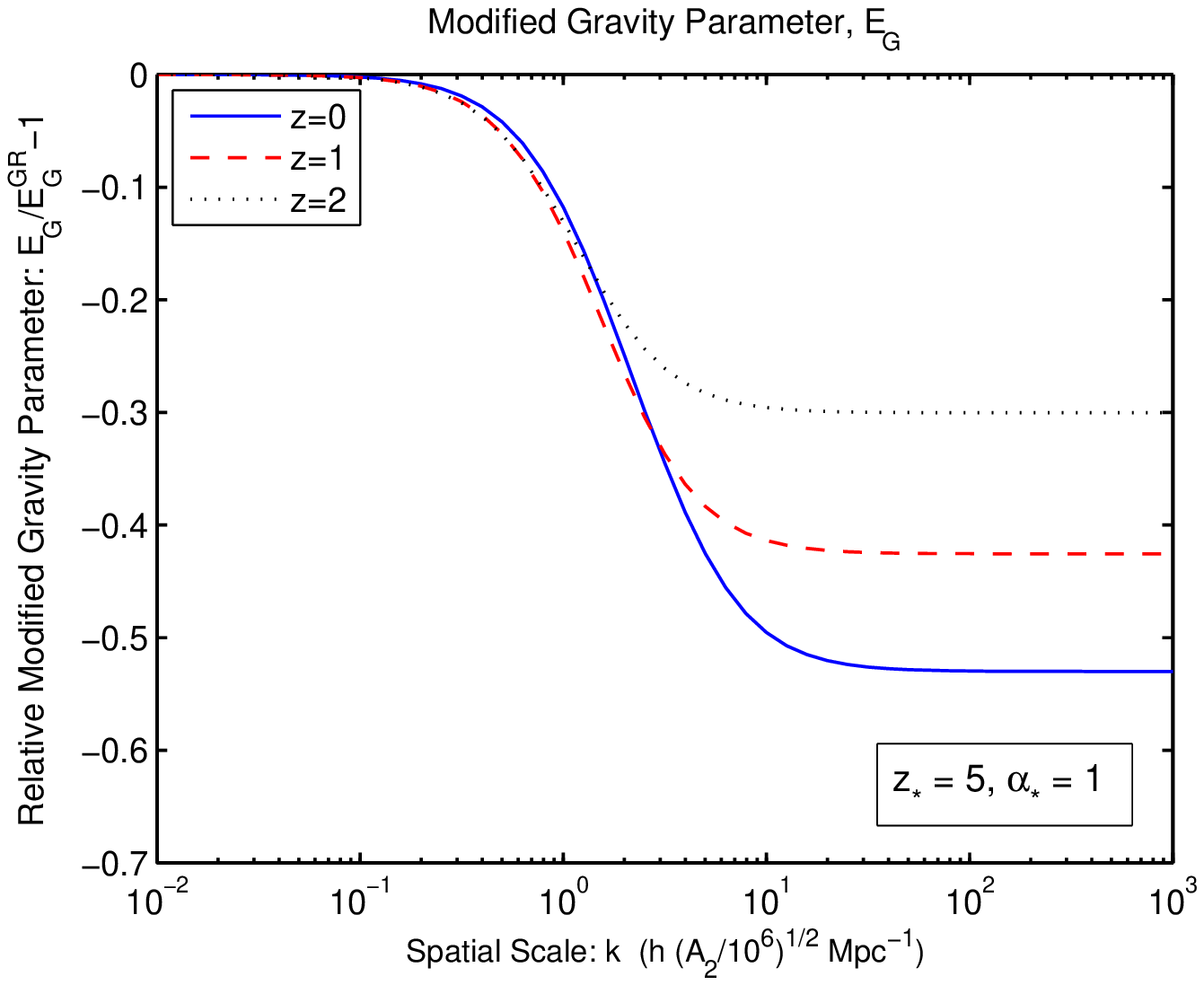}
\caption[]{Effect of the symmetron field on large scale structure formation.  From top to bottom the plots show: the predicted deviation of the (linear bias corrected) galaxy growth, $f_{\rm gal}^{\rm bc}$, from its GR value, $f_{\rm GR}$, the predicted value of the slip parameter, $\Sigma_{\rm \kappa m}$, extrapolated from weak lensing measurements, the predicted slip parameter extrapolated from ISW measurements, $\Sigma_{\rm I m}$, and the relative deviation of the modified gravity parameter, $E_{\rm G}$, from its GR value. These plots are for $[\alpha_{\star} = 1,\,\, z_{\star} = 5]$, and show the values of $f_{\rm gal}^{\rm bc}$, $\Sigma_{\rm \kappa m}$ and $\Sigma_{\rm Im}$ at the present day, $z=0$ (solid blue line), $z=1$ (dashed red line) and $z=2$ (dotted black line) for different values of the inverse spatial scale, $k$.
\label{fig2}}
\end{center}
\end{figure*}
Figs.~\ref{fig1} and \ref{fig2} show $f_{\rm gal}$, $\Sigma_{\kappa m}$, $\Sigma_{\kappa I}$ and $E_{\rm G}$ with $z_{\star} = 2$ and $z_{\star} = 5$ respectively. In both figures we have taken $\alpha_{\star}=1$ and $\Omega_{\rm m0} =0.27$.  Local tests of gravity only constrain $A_2 \gtrsim 10^{5}\textrm{--} 10^{6}$ which sets the spatial scale below which the modification of gravity is apparent. It is clear from the figures that there are regions of the allowed parameter space where the modifications to gravity are potentially large.

However, with $A_2 \approx 10^{6}$, modification to structure formation occur on and below Mpc scales, and on these scales the linear perturbation approximation breaks down at late times.  As with the environmentally dependent dilaton model \cite{bbds2010,bbdls2011}, we expect the onset of nonlinear structure to suppress the fifth force giving a weaker observational signal, and this is confirmed in \cite{wmld2011}. However, unlike the dilaton model, the  redshift at which gravity deviates from GR, $z_{\star}$, is a free parameter and if this occurs early enough then structure will still be linear on Mpc scales and the linearised approximation will hold and non-negligible changes to structure formation could occur. We will consider possible observational constraints in the future.

\section{Summary and Conclusions}

\label{sect:con}

In this paper we have studied the background cosmology and the evolution of 
linear perturbations for the symmetron model. The model utilises the so-called symmetron mechanism to suppress the fifth force mediated by the scalar symmetron field in regions with high matter density to evade the local gravity constraints. In the regions with low matter density, the fifth force is unsuppressed, which could have significant consequences on the cosmic structure formation.

The symmetron mechanism works because the effective potential $V_{\rm eff}(\phi)$, which governs the evolution of $\phi$, depends explicitly on the local matter density. If the matter density is high, the potential has a global minimum at $\phi=0$ about which it is symmetric; then because the coupling strength of the symmetron field to matter is proportional to $\phi$, the coupling vanishes and so does the fifth force. When the matter density drops below some critical value $\rho_\star$, the symmetry of the potential is broken and the minimum of $V_{\rm eff}$ moves aways from $\phi=0$; because the scalar field tries to track that minimum, the coupling becomes nonzero and so does the fifth force. When inhomogeneous matter distribution is considered, it is possible that the symmetry in $V_{\rm eff}$ is broken in a region, but later the matter density in that region exceeds $\rho_\star$ again because of the structure formation process, and the symmetry is restored so that the fifth force vanishes again. Obviously such a phenomenon is quite nontrivial to study.

We have studied the behaviour of the symmetron field in background cosmology, when the symmetry breaking happens in the matter dominated era. We have found that immediately after the symmetry breaking, the symmetron field lags behind its effective potential minimum and therefore has a negative mass-squared. This implies that it will experience a tachyonic period during which both the field and its perturbation grow exponentially. However this period is fairly short and the symmetron field will then quickly settle down to the (moving) minimum of $V_{\rm eff}$, where it oscillates. We find analytical solutions to these different phases and show that they agree quite well with numerical results.

We have then investigated the general behaviour of the linear perturbations in the symmetron model. The symmetron perturbations have an effective mass-squared $k^2/a^2+m^2(\phi)$, in which $k$ is the wavenumber of the perturbation and $m(\phi)$ is the mass of the background symmetron field. On very small scales, $k^2/a^2+m^2(\phi)>0$ even if $m^2<0$, and there is no tachyonic instability. In this case we notice that the growth of matter perturbation is governed by an enhanced Newton constant which is $G=(1+\alpha)G_N$ when the size of the perturbation is smaller than the Compton radius of the symmetron field (which means that the perturbation is within the reach of the fifth force), while it is the same as in GR if the perturbation is bigger than the Compton radius.

On larger scales it is possible that $k^2/a^2+m^2(\phi)<0$ given that $m^2<0$ and the symmetron perturbation will experience the tachyonic growth, which naively could be a problem as it would mean that everything blows up. We find, however, that the tachyonic behaviour does not affect the growth of matter perturbation significantly, at least on sub-horizon scales. The reason is that the tachyonic period is so short that the even faster growth of $\phi$ itself has negligible effect in cosmology, let alone that of $\delta\phi$. This conclusion has been confirmed by our numerical results.

Finally, we have studied the structure formation in the linear regime under the adiabatic approximation, i.e., assuming that the scalar field always tracks the minimum of its effective potential. Our result shows that, depending on the redshift $z_\star$ of the symmetry breaking, the effect of the fifth force on the large-scale structure could be important. If $z_\star$ is big, then the symmetry breaking happens at earlier times when the matter perturbations are well described by the linear theory and our analysis gives accurate result. In such cases we find that the fifth force comes into effect fairly early and we expect bigger deviations from the $\Lambda$CDM model predictions. If $z_\star$ is small, then the fifth force takes effect quite late, when matter perturbations have already entered the nonlinear regime. In such cases the linear theory is no longer accurate and better results can only be obtained using nonlinear numerical simulations \cite{wmld2011}.

To summarise, the symmetron mechanism is shown to work very successfully to suppress the fifth force to an undetectable level in our solar system, while still allow non-negligible effects to be found in the cosmological structure formation process. This is very important for the study of dark energy, because it means that one has to look at the small scales (relevant for galaxies and galaxy clusters) to find the implications of a theory which is supposed to modify cosmology on the very largest scales. Full constraint of the symmetron theory parameters will be studied in the future.

\

\

{\it Notes added}: After this work had been completed, we noticed
Ref.~\cite{hklm2011} on arXiv which also contained an
investigation of the background cosmology in the symmetron model.
Our work mainly focus on the linear perturbation growth in the
model, and is thus different from Ref.~\cite{hklm2011}. A separate
work, Ref.~\cite{wmld2011}, concentrates on the structure
formation of the symmetron model in the nonlinear regime.

\begin{acknowledgments}
This work is supported in part by STFC (CvdB, ACD, BL). 
BL is supported by the Queens' College and Department of Applied Mathematics
and Theoretical Physics at the University of Cambridge.
\end{acknowledgments}


\begin{thebibliography}
\bibitem{} \ifx\csname natexlab\endcsname\relax \fi \expandafter\ifx\csname
bibnamefont\endcsname\relax

\fi \expandafter\ifx\csname bibfnamefont\endcsname\relax

\fi \expandafter\ifx\csname citenamefont\endcsname\relax

\fi \expandafter\ifx\csname url\endcsname\relax

\fi \expandafter\ifx\csname urlprefix\endcsname\relax

\fi \providecommand{\bibinfo}[2]{#2}
\providecommand{\eprint}[2][]{\url{#2}}

\bibitem{cst2006} E.~J.~Copeland, M.~Sami and S.~Tsujikawa, Int.~J.~Mod.~Phys.~D{\bf 15}, 1573 (2006).

\bibitem[\protect\citeauthoryear{Wang {\it et al}}{2000}]{wcos2000} L.~Wang, R.~R.~Caldwell, J.~P.~Ostriker and P.~Z.~Steinhardt, Astrophys.~J., {\bf 530}, 17 (2000).

\bibitem[\protect\citeauthoryear{Armendariz-Picon {\it et al}}{2000}]{ams2000} C.~Armendariz-Picon, V.~Mokhanov and P.~J.~Steinhardt, Phys.~Rev.~Lett., {\bf 85}, 854438 (2000).

\bibitem{Wetterich:1987fm}
  C.~Wetterich,
  Nucl.\ Phys.\  {\bf B302 } (1988)  668.

\bibitem[\protect\citeauthoryear{Amendola}{2000}]{a2000} L.~Amendola, Phys.~Rev.~D{\bf62}, 043511 (2000).

\bibitem[\protect\citeauthoryear{Perrotta \& Baccigalupi}{1999}]{pb1999} F.~Perrotta and C.~Baccigalupi, Phys.~Rev.~D{\bf61}, 023507 (1999).

\bibitem{kw2004a} J.~Khoury and A.~Weltman, Phys.~Rev.~Lett.~{\bf 93}, 171104 (2004).

\bibitem{kw2004b} J.~Khoury and A.~Weltman, Phys.~Rev.~D{\bf 69}, 044026 (2004).

\bibitem{Brax:2004qh}

Ph.~ Brax, C.~ van de Bruck, A-C~ Davis, J.~ Khoury and  A.~Weltman,
Phys.Rev.D70:123518,2004
[arXiv: astro-ph/0408415]

\bibitem{ms2006} D.~F.~Mota and D.~J.~Shaw, Phys.~Rev.~Lett.~{\bf 97}, 151102 (2006).

\bibitem{ms2007} D.~F.~Mota and D.~J.~Shaw, Phys.~Rev.~D{\bf 75}, 063501 (2007).

\bibitem{lb2007} B.~Li and J.~D.~Barrow, Phys.~Rev.~D{\bf 75}, 084010 (2007).

\bibitem{bbds2008} P.~Brax, C.~Van~de~Bruck, A.-C.~Davis and D.~J.~Shaw, Phys.~Rev.~D{\bf 78}, 104021 (2008).

\bibitem{bbds2010} P.~Brax, C.~Van~de~Bruck, A.-C.~Davis and D.~J.~Shaw (2010), arXiv:1005.3735 [astro-ph.CO].

\bibitem{bbdls2011} P.~Brax, C.~Van~de~Bruck, A.-C.~Davis, B.~Li and D.~J.~Shaw, Phys.~Rev.~D{\bf 83}, 104026 (2011).

\bibitem{dgp2000} G.~Dvali, G.~Gabadadze and M.~Porrati, Phys.~Lett.~B{\bf 485}, 208 (2000).

\bibitem{nrt2009} A.~Nicolis, R.~Rattazzi and E.~Trincherini, Phys.~Rev.~D{\bf 79}, 064306 (2009).

\bibitem{dev2009} C.~Deffayet, G.~Esposito-Farese and A.~Vikman, Phys.~Rev.~D{\bf 79}, 084003 (2009).

\bibitem{dde2009} C.~Deffayet, S.~Deser and G.~Esposito-Farese, Phys.~Rev.~D{\bf 80}, 064015 (2009).

\bibitem{dt2010} A.~De~Felice and S.~Tsujikawa, Phys.~Rev,~Lett., {\bf 105}, 111301 (2010).

\bibitem[Hinterbichler \& Khoury (2010)]{hk2010} K.~Hinterbichler and J.~Khoury, Phys.~Rev.~Lett.~{\bf 104}, 231301 (2010).

\bibitem{Bertotti:2003rm}
  B.~Bertotti, L.~Iess, P.~Tortora,
  Nature {\bf 425 } (2003)  374.


\bibitem{Brax:2010ai}
  P.~Brax, J.~-F.~Dufaux, S.~Mariadassou,
  Phys.\ Rev.\  {\bf D83 } (2011)  103510.
  [arXiv:1012.4656 [hep-th]].

\bibitem[Hinterbichler~{\it et~al.} (2011)]{hklm2011} K.~Hinterbichler, J.~Khoury, A.~Levy and A.~Matas (2011), arXiv:1107.2112 [astro-ph.CO].

\bibitem{Brax:2009ab}
  P.~Brax, C.~van de Bruck, A.~-C.~Davis, D.~Shaw,
  JCAP {\bf 1004 } (2010)  032.

\bibitem[Winther~{\it et~al.} (2011)]{wmld2011} A.~-C.~Davis, B.~Li, David~F.~Mota and H.~A.~Winther (2011), arXiv:1108.xxxx [astro-ph.CO].

\end{thebibliography}
\end{document}